\documentclass[11pt]{article}
\pdfoutput=1
\usepackage{amsmath,amssymb,amsfonts,graphicx,slashed,color,cite}%, epsf, dsfont, dcolumn,yfonts}
\usepackage[colorlinks=true,linkcolor=blue, citecolor=blue, linktoc=all]{hyper ref}

\DeclareMathAlphabet{\mathbbold}{U}{bbold}{m}{n}

\parskip=\baselineskip

\newcommand{\myfig}[3]{
	\begin{figure}[ht]
	\centering
	\includegraphics[width=#2cm]{#1}\caption{#3}\label{fig:#1}
	\end{figure}
	}

\newcommand{\cDsl}{{{\cal D}\kern-.65em /\,}}
\newcommand{\cDslsm}{{{\cal D}\kern-.5em /\,}}
\newcommand{\nabslsm}{\nabla\kern-.55em /}
\newcommand{\pasl}{\pa\kern-.55em /}
\newcommand{\psl}{p\kern-.45em /}
\newcommand{\Dsl}{D\kern-.65em /}
\newcommand{\Asl}{A\kern-.55em /}
\newcommand{\nabsl}{\nabla\kern-.65em /\kern+.2em}
\newcommand{\qsl}{q\kern-.5em /}
\newcommand{\ksl}{k\kern-.5em /}
\newcommand{\rsl}{r\kern-.5em /}

\newcommand{\cDslLCsq}{{\stackrel{\circ}{\cDsl^{\kern2pt 2}}}}

\newcommand{\cL}{\mathcal{L}}

\newcommand\cc[1]{#1^{^{\kern-6pt \circ}}\kern2pt}

\newcommand\abs[1]{\big|#1\big|} %puts argument in abs value

\newcommand{\pa}{\partial}

\newcommand{\beq}{\begin{equation}}
\newcommand{\eeq}{\end{equation}}
\newcommand{\beqn}{\begin{eqnarray}}
\newcommand{\eeqn}{\end{eqnarray}}

\newcommand{\mTr}{\mathrm{Tr}}

\def\dalemb#1#2{{\vbox{\hrule height .#2pt
\hbox{\vrule width.#2pt height#1pt \kern#1pt
\vrule width.#2pt}
\hrule height.#2pt}}}

\newcommand{\cH}{\mathcal{H}}
\newcommand{\cS}{\mathcal{S}}
\newcommand{\cT}{\mathcal{T}}
\newcommand{\bs}{\boldsymbol}
\newcommand{\relent}[2]{S\left(#1\abs{}#2\right)} %relative entropy of two states

\begin{document}

%\begin{center}
%\title
%\end{center}
%\centerline{[Draft Version of \today]}
%\vskip 2 cm
%\centerline{{\bf Good people}}

%\vspace{.5cm}
%\centerline{\it \uiucaddress}

%\centerline{\it \upennaddress}
%\vspace{2cm}
\title{Multi-Boundary Entanglement in \\ Chern-Simons Theory and Link Invariants}
\author{
{Vijay Balasubramanian$^{1,2}$, Jackson R. Fliss$^3$, Robert G. Leigh$^3$ and Onkar Parrikar$^1$}\\\\
{$^1$\small{ \emph{David Rittenhouse Laboratory, University of Pennsylvania, 209 S.33rd Street, Philadelphia PA, 19104, U.S.A.}} }\\
{$^2$ \small{ \emph{Theoretische Natuurkunde, Vrije Universiteit Brussel (VUB), and }}}\\
{\small{\emph{International Solvay Institutes, Pleinlaan 2, B-1050 Brussels, Belgium}}}\\
{$^3$\small{ \emph{Department of Physics, University of Illinois, 1110 W. Green Street, Urbana IL, 61801, U.S.A. }}}
}
\maketitle
\begin{abstract}
We consider Chern-Simons theory for gauge group $G$ at level $k$ on 3-manifolds $M_n$ with boundary consisting of $n$ topologically linked tori.   The Euclidean path integral on  $M_n$ defines a quantum state on the boundary, in the $n$-fold tensor product of the torus Hilbert space. We focus on the case where $M_n$ is the link-complement of some $n$-component link inside the three-sphere $S^3$. The entanglement entropies of the resulting states define framing-independent link invariants which are sensitive to the topology of the chosen link. 
%of the chosen link.
   For the Abelian theory at level $k$ ($G= U(1)_k$) we give a general formula for the entanglement entropy associated to an arbitrary $(m|n-m)$ partition of a generic $n$-component link into sub-links. The formula involves the number of solutions to certain Diophantine equations with coefficients related to the Gauss linking numbers (mod $k$) between the two sublinks.  This formula connects simple concepts in quantum information theory, knot theory, and number theory, and shows that entanglement entropy between sublinks vanishes if and only if they have zero Gauss linking (mod $k$).   For $G = SU(2)_k$, we study various two and three component links.  We show that the 2-component Hopf link is maximally entangled, and hence analogous to a Bell pair, and that the Whitehead link, which has zero Gauss linking, nevertheless has entanglement entropy. Finally, we show that the Borromean rings have a ``W-like" entanglement  structure (i.e., tracing out one torus does {\it not} lead to a separable state), and give examples of other 3-component links which have ``GHZ-like'' entanglement (i.e., tracing out one torus {\it does} lead to a separable state).
\end{abstract}

\pagebreak

\section{Introduction}

An important open question in quantum mechanics and quantum information theory is to understand the possible patterns of entanglement that can  arise naturally in field theory.    The local structure of wavefunctions is typically determined largely by the locality of physical Hamiltonians because interactions create entanglement.  However, entanglement is a global property and very little is known about how it can be organized over long distances.   One way of thinking about this is to consider multiple disjoint regions that are sufficiently separated so that locality by itself will not prescribe the structure of entanglement.   A challenge is that there is no general prescription for even classifying the patterns of entanglement between multiple disjoint entities.   For three qubits, up to local operations, or more precisely up to SLOCC (Stochastic Local Operations and Classical Communication) transformations of the state, there are precisely two non-trivial classes of multipartite entanglement \cite{Dur:2000zz} -- the GHZ class, represented by the state $(|111\rangle + |000\rangle)/\sqrt{2}$, has the property that tracing over one qubit disentangles the state, while in the W class, represented by $(|100\rangle + |010\rangle + |001\rangle)/\sqrt{3}$, a partial trace still leaves an entangled state of two qubits.  A similar analysis of entanglement classes is not known in general for $n$ qubits, or in the more physical case of LOCC equivalence, let alone for disjoint regions of a field theory.

Recently the AdS/CFT correspondence was proposed as a tool for studying multi-partitite etanglement.  The authors of \cite{Balasubramanian:2014hda,Marolf:2015vma} examined the multi-boundary three-dimensional wormhole solutions of \cite{Brill:1995jv,Aminneborg:1997pz,Brill:1998pr,Aminneborg:1998si,Krasnov:2000zq,Krasnov:2003ye,Skenderis:2009ju} and found non-trivial entanglement, computed through the holographic Ryu-Takayangi formula \cite{Ryu:2006bv},  between subsets of boundary components.   One interesting result was that although there were regions of parameter space where the entanglement between boundaries was entirely multi-partite, it was never of the GHZ type.   In special limits it was also possible to analyze the structure of the CFT wavefunction in terms of the OPE coefficients.  However, it was difficult to carry out a computation of entanglement entropies in the field theory at a generic point in the parameter space. 

While the field theory calculation of multi-boundary entanglement entropies is difficult in general, one simple case where this can be done is in a topological quantum field theory \cite{Witten:1988hf, Witten:1988ze, Atiyah1988} defined on a manifold $M_n$, with boundary $\Sigma_n$ consisting of a union of $n$ disjoint components $\{ \sigma_1, \sigma_2, \cdots \sigma_n \}$.     The Euclidean path integral for this theory as a functional of data on the boundary defines a wavefunction on $\Sigma_n$.   This wavefunction is defined on the tensor product of Hilbert spaces ${\cal H}_i$ associated with the different boundary components.   Because the theory is topological there will be no local dynamics, and all of the entanglement arises from the topological properties of $M_n$.  This allows us to focus attention on global features of entanglement, and we can hope that geometric and topological tools will come to our aid.

Here, we explore these ideas in the context of Chern-Simons gauge theories in three dimensions (see \cite{Witten:1988hf, Marino:2004uf} and references there-in).  Bi-partite entanglement of connected spatial sections in such theories was studied in \cite{Kitaev:2005dm, PhysRevLett.96.110405, Dong:2008ft}.   By contrast, we consider Chern-Simons theory for group $G$ at level $k$ defined on 3-manifolds $M_n$ with disconnected boundaries, namely $n$ linked tori. More precisely, we will choose $M_n$ to be link complements (see definition below) of $n$-component links in $S^3$; the wavefunctions on the tori in this case can be explicitly written in terms of \emph{coloured link invariants}.  For $G=U(1)_k$ this leads to a general formula for the entanglement entropy of any bipartition of the link into sub-links.  Further, the entropy vanishes if and only if the Gauss linking number vanishes (modulo $k$)between the sub-links in the bipartition.  It is also possible to construct states with non-zero tripartite mutual information of both signs.  For $G=SU(2)_k$ we explicitly calculate entanglement entropies for a variety of 2- and 3-component links, and show that: (a) the Hopf link is the analog of a maximally-entangled Bell pair, (b) while the $U(1)$ entanglement is only sensitive to the Gauss linking number, the non-Abelian entanglement also detects more subtle forms of topology, and (c)  GHZ-like states and W-like states are both realizable in terms of links with different topologies.  Overall, multi-boundary entanglement entropy in Chern-Simons theory computes a framing-independent link invariant with physical motivation, and hence gives a potentially powerful tool for studying knots and links. Additionally, this setup also gives a calculable arena for the study of multi-partite entanglement.

%For this reason, the topological field theory computations of this paper provide a useful a counterpoint. 
Interestingly, at the classical level the three-dimensional theories of gravity studied in the holographic approach to multi-partite entanglement \cite{Balasubramanian:2014hda,Marolf:2015vma}  can themselves be written as Chern-Simons theories of the group $SL(2,R) \times SL(2,R)$.   While it is not clear that 3d quantum gravity is entirely described by Chern-Simons theory \cite{Maloney:2007ud}, it is intriguing to speculate that we could use our Chern-Simons techniques to directly compute entanglement in three dimensional gravity. 

The rest of the paper is organized as follows: in Section \ref{sec2}, we will construct the multi-boundary states we are interested in, and review some concepts required for later calculations. In Section \ref{sec3}, we will consider Chern-Simons theory for $G=U(1)_k$, and compute the entanglement entropy for a bi-partition of a generic $n$-component link into sub-links. In Section \ref{sec4}, we will consider multi-boundary entanglement in $G=SU(2)_k$ Chern-Simons. Here we will study several examples of two and three-component links and try to extract general lessons from these examples. Finally, we end with a discussion of open questions and future work in Section \ref{sec5}.

\section{Multi-boundary States in Chern-Simons theory}\label{sec2}
%\subsection{Generalities}

%\begin{itemize}
%\item review $\cS, \cT$; define in CFT, then explain presence in surgery (note $su(2)_k$ S-matrix derived in appendix)
%\item put in discussion of surgery, so that later calcs are more straightforward.
%\item Previous studies of entanglement in continuum Chern-Simons theories\cite{Dong:2008ft} focussed on bi-partite entanglement in the context of connected spatial sections. Here, we formulate entanglement for multi-component spatial sections, corresponding to C-S on link complements.
%\item put in conclusion: could redo calculations for other CFTs -- need to know S-matrices, etc. (see \cite{Dong:2008ft})
%
%\end{itemize}

We consider Chern-Simons theory with gauge group $G$ at level $k$. The action of the theory on a  3-manifold $M$ is given by
\beq
S_{CS}[A] = \frac{k}{4\pi} \int_{M} \mathrm{Tr}\,\left(A\wedge dA + \frac{2}{3} A \wedge A \wedge A\right),
\eeq
where $A= A_{\mu}dx^{\mu}$ is a gauge field (or equivalently, a connection on a priniple $G$-bundle over $M$). The equation of motion corresponding to the above action is 
\beq
F = dA+A\wedge A =  0.
\eeq
Since the equation of motion restricts the phase space to flat connections (modulo gauge transformations), the only non-trivial, gauge invariant operators in the theory are \emph{Wilson lines} along non-contractible cycles in $M$:
\beq
W_{R}(L)  = \mathrm{Tr}_{R}\, \mathcal{P}\,e^{i\oint_L A},
\eeq
where $R$ is a representation of $G$, $L$ is an oriented, non-contractible cycle in $M$ and the symbol $\mathcal{P}$ stands for path-ordering along the cycle $L$. If $M$ has a boundary $\Sigma$, then the path-integral of the theory on $M$ with Wilson line insertions,  and boundary conditions $A|_{\Sigma} = A^{(0)}$ imposed on $\Sigma$,\footnote{When $M$ has a boundary, then the action must be augmented by including certain boundary terms, which correspond to picking a Lagrangian submanifold in phase space. We will not need to dwell on these details in the present paper.} namely
\beq
\Psi_{(R_1,L_1),\cdots, (R_n, L_n)}[ A^{(0)}] = \int_{A|_{\Sigma} = A^{(0)}} [DA] e^{iS_{CS}[A]} \,W_{R_1}(L_1) \cdots W_{R_n}(L_n)
\eeq
is interpreted as the wavefunction of a state in the Hilbert space $\mathcal{H}(\Sigma; G, k)$ which Chern-Simons theory associates to $\Sigma$. In this paper, we consider states in the $n$-fold tensor product $\cH^{\otimes n}$, where $\cH=\cH(T^2;G,k)$ is the Hilbert space of Chern-Simons theory for the group $G$ at level $k$ on a torus.  These states can be understood as being defined on $n$ copies of $T^2$, namely the spatial manifold $\Sigma_n$ \
\beq
\Sigma_n = \amalg_{i=1}^n T^2, 
\eeq
where $\amalg$ denotes disjoint union (see Figure \ref{fig:fig1.pdf}).
%  Which state in particular do we want to consider? 
A natural way to construct states in a QFT is by performing the Euclidean path integral of the theory on a 3-manifold $M_n$ whose boundary is $\pa M_n = \Sigma_n$.      In a general field theory the state constructed in this way will depend on the detailed  geometry of $M_n$, for instance the choice of metric on $M_n$, but in our situation only the topology of $M_n$ matters.  However, there are many topologically distinct Euclidean  3-manifolds with the same boundary, and the path integrals on these manifolds will construct different states on $\Sigma_n$.    We will focus on a simple class of such 3-manifolds, which we will now describe.

%We consider states in the $n$-fold tensor product $\cH^{\otimes n}$ of the Hilbert space $\cH(T^2;G,k)$ (or simply $\cH$ for short) of Chern-Simons theory for the group $G$ at level $k$ on a torus. From standard rules of quantum field theory, these states can be understood as being defined on $n$ copies of $T^2$, namely the spatial manifold $\Sigma_n$ given by 
%\beq
%\Sigma_n = \amalg_{i=1}^n T^2, 
%\eeq
%where $\amalg$ denotes disjoint union (see Figure \ref{fig:fig1.pdf}). Which state in particular do we want to consider? A natural way to construct states in a QFT is by performing the path integral of the theory on a 3-manifold $M_n$ whose boundary $\pa M_n = \Sigma_n$. Now given that $\Sigma_n$ is the disjoint union of $n$ tori, it is clear that the choice of $M_n$ is far from unique; there are multiple topologically distinct manifolds $M_n$ we can pick, each corresponding to a different state on $\Sigma_n$. Furthermore, in a general QFT the state also depends on the details of the geometry of $M_n$, such as for instance the choice of metric on $M_n$. However, since we are here interested in a topological quantum field theory, we need only worry about topologically distinct choices. In this situation, there is a simple way to construct the 3-manifolds $M_n$ described above, which we will now explain. 

\myfig{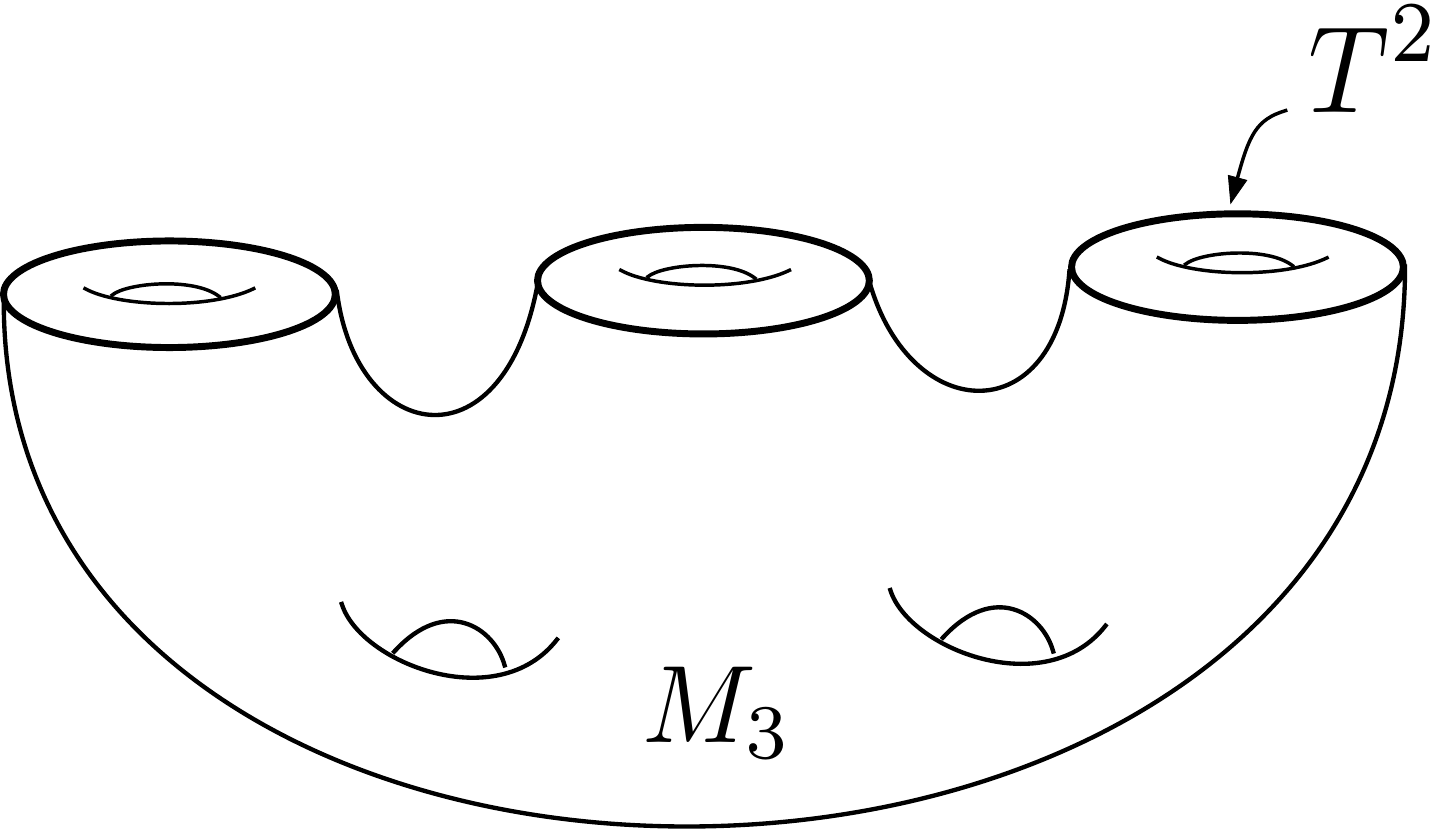}{8}{\small{\textsf{The spatial manifold $\Sigma_n$ for $n=3$ is the disjoint union of three tori. $M_n$ is a 3-manifold such that $\pa M_n = \Sigma_n$.}}}

We start with a connected, closed 3-manifold (i.e., a connected, compact 3-manifold without boundary) $X$. An $n$-component \emph{link} in $X$ is an embedding of $n$ (non-intersecting) circles in $X$. (Note that 1-component links are conventionally called \emph{knots}.) We will sometimes use Rolfsen notation to denote a link $\cL$ as $\cL=c^n_m$, where $c$ is the number of crossings, $n$ is the number of components in the link, and $m$ is the chronological rank at which the link is presented in the Rolfsen table \cite{rolfsen1976knots} for a given $c$ and $n$. We will sometimes merely denote a generic $n$-component link as $\cL^n$, when we do not need to choose a particular link. We will label the $n$ circles which constitute the link as $L_1, \ldots, L_n$, so $\cL^n = L_1 \cup L_2 \cup \cdots \cup L_n$. Now in order to construct the desired 3-manifold $M_n$, we pick a link $\cL^n$ in $X$ and \emph{drill} out a tubular neighbourhood $\tilde{\cL}^n$ of the link in $S^3$. In other words, we take $M_n$ to be the complement of $\cL^n$ in $X$, i.e., $M_n = X - \tilde{\cL}^n$ (see Figure \ref{fig:fig2.pdf}). This is a standard construction; the 3-manifold $M_n$ we have obtained starting from $X$ and $\cL^n$ is called the \emph{link complement} of $\cL^n$ in $X$. Since $\cL^n$ is an $n$-component link, its link complement $M_n$ is a manifold with precisely the desired boundary 
\beq
\pa M_n = \amalg_{i=1}^n T^2.
\eeq
We can therefore perform the path-integral of Chern-Simons theory on $M_n$, and obtain a state on $\Sigma_n$. In fact, \emph{every topological 3-manifold $M_n$ which has the disjoint union of $n$ tori as its boundary, is a link-complement $X - \cL^n$, for some closed 3-manifold $X$ and an $n$-component link $\cL^n$ in $X$.} This construction assigns a state $\left| \cL^n, X\right \rangle $ to every pair $(X, \cL^n)$ -- we will sometimes refer to these states as \emph{link states}. In this paper, we will focus on the class of states constructed this way, but where we take $X$ to be the 3-sphere $S^3$.

\myfig{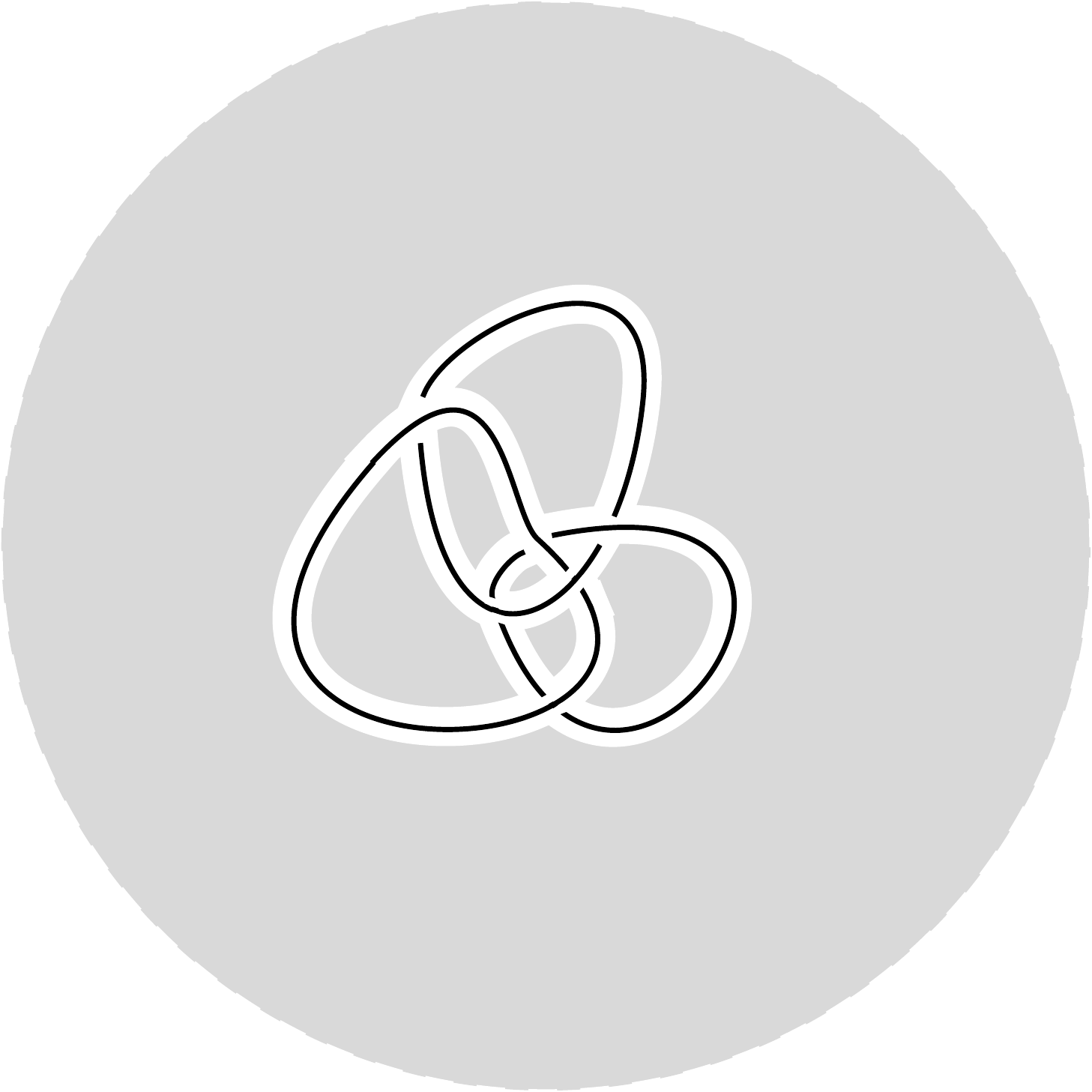}{5.5}{\small{\textsf{The link complement (the shaded region) of a 3-component link (bold lines) inside the three-sphere. The white region indicates a tubular neighbourhood of the link which has been drilled out of the 3-sphere.}}}

To further understand the state $\left| \cL^n, S^3\right \rangle $, or simply $\left| \cL^n\right \rangle $ for short, we need to know some details about the Hilbert space of Chern-Simons theory on a torus $T^2$ \cite{Witten:1988hf}. Let us picture the 2-torus as the boundary of a solid torus inside $S^3$ (see Figure \ref{fig:fig3.pdf}). We pick two simple cycles on the torus which generate its fundamental group and label them $\bs{m}$ and $\bs{\ell}$, with $\bs{m}$ being the \emph{meridian}, i.e., contractible inside the solid torus. The choice of $\bs{\ell}$, called the \emph{longitude}, is not unique. But let us make the canonical choice for $\bs\ell$, namely the one which is contractible in the complement of the torus inside $S^3$; we will later return to this point, which is related to \emph{framing}. In order to construct a basis for the Hilbert space $\cH(T^2;G,k)$ we perform the Chern-Simons path integral on the solid torus with a Wilson line in the representation $R_j$ placed in the bulk of the solid torus running parallel to the longitude cycle $\bs{\ell}$, where the index $j$ denotes an \emph{integrable} representation of the gauge group $G$ at level $k$. This gives a state on $T^2$ which we call $|j\rangle$. The conjugate of this state $\langle j|$ can be thought of in terms of the path integral on the solid torus with a Wilson line in the conjugate representation $R_j^*$. By letting $j$ run over all the integrable representations \cite{Gepner:1986wi} of $G$, we obtain a basis for the torus Hilbert space. Notably, the Hilbert space $\mathcal{H}$ obtained in this way is finite dimensional. For example if we take $G=SU(2)_k$, the integrable representations are labelled by their spin $j$ for $  j =0, \frac{1}{2}, \cdots, \frac{k}{2}$, and so $\mathrm{dim}(\mathcal{H}\left(T^2; SU(2), k)\right) = k+1$. Similarly in $G=U(1)_k$, the allowed representations are labeled by integer-valued charges $0 \leq q <k$, and so $\mathrm{dim}\left(\mathcal{H}(T^2; U(1), k)\right) = k$. We also note that the modular group $SL(2,\mathbb{Z})$ of large diffeomorphisms of the torus, generated by
\beq
\mathcal{T} : \tau \to \tau + 1,\;\;\; \cS : \tau \to -\frac{1}{\tau}
\eeq
acts naturally on $\mathcal{H}(T^2; G, k)$. For example in the $U(1)_k$ theory, these operators take the following simple form \cite{Dong:2008ft} in the basis we introduced above\footnote{The $\cT$ matrices generally also contain an additional overall phase proportional to the central charge; we have omitted this phase above since it will not play any role in our discussion.}:
\beq
\mathcal{T}_{q_1,q_2} = e^{2\pi i h_{q_1} } \delta_{q_1,q_2},\;\;\; \cS_{q_1,q_2} = \frac{1}{\sqrt{k}} e^{\frac{2\pi i q_1 q_2}{k}}
\eeq
where $h_{q} = q^2/2k$. Similarly, for $SU(2)_k$ we have
\beq
\mathcal{T}_{j_1,j_2} = e^{2\pi i h_{j_1}} \delta_{j_1,j_2},\;\;\; \cS_{j_1,j_2} = \sqrt{\frac{2}{k+2}}\sin\left(\frac{\pi (2j_1+1)(2j_2+1)}{k+2}\right)
\eeq
where $h_{j} = \frac{j(j+1)}{k+2}$. It is not hard to check that these matrices satisfy the relations $\cS^2 = 1$ and $(\cS \cT)^3 = 1$. 

\myfig{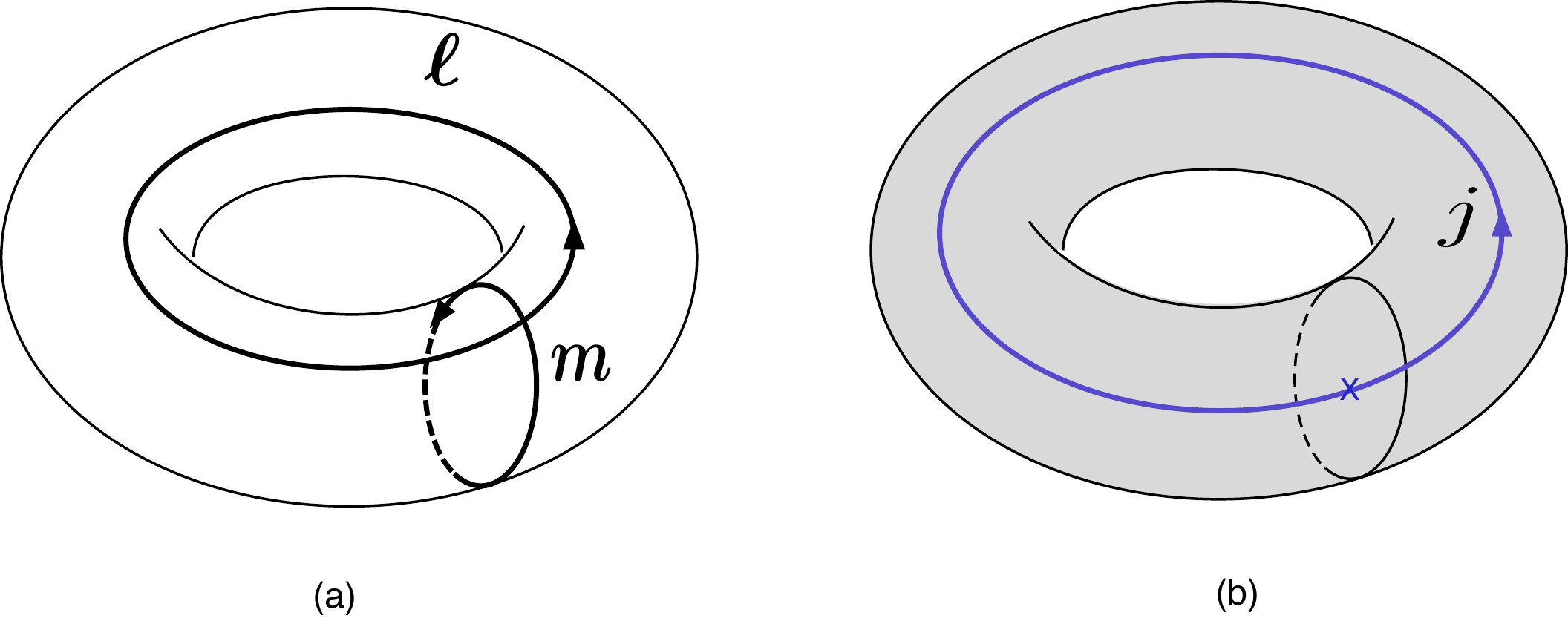}{10}{\small{\textsf{(a) The meridian and longitude cycles on a torus $T^2$. (b) The state $|j\rangle$ corresponds to a Wilson line in the representation $j$ placed in the bulk of the solid torus.}}}

Now let us write the state $|\cL^n\rangle \in \mathcal{H}^{\otimes n}$ obtained by performing the path-integral of Chern-Simons theory on the link complement of the link $\cL^n$ in terms of the above basis vectors:
\beq
|\cL^n\rangle = \sum_{j_1,\cdots, j_n} C_{\cL^n}(j_1, j_2, \cdots j_n) |j_1,j_2, \cdots, j_n\rangle,\qquad  |j_1,j_2, \cdots, j_n\rangle\equiv |j_1\rangle \otimes |j_2\rangle \otimes |j_n\rangle
\eeq
where $C_{\cL^n}(j_1,\cdots, j_n)$ are complex coefficients, which we can write explicitly as
\beq
C_{\cL^n}(j_1, j_2, \cdots j_n)= \langle j_1,j_2,\cdots j_n | \cL^n\rangle  \, .
\eeq
%From the general rules of quantum field theory, this precisely 
Operationally, this corresponds to gluing in solid tori along the boundary of the link complement $S^3-\cL^n$, but with Wilson lines in the representation $R^*_{j_i}$ placed in the bulk of the $i^{th}$ torus.  Thus, the coefficients $C_{\cL^n}(j_1,\cdots j_n)$ are precisely the \emph{coloured link invariants} of Chern-Simons theory with the representation $R^*_{j_i}$ placed along the $i^{th}$ component of the link:
\beq
C_{\cL^n}(j_1, \cdots, j_n) = \left\langle W_{R^*_{j_1}}(L_1) \cdots W_{R^*_{j_n}}(L_n)\right\rangle_{S^3} \label{cli} \, ,
\eeq
where we recall that $L_i$ are the individual circles which constitute the link, namely $\cL^n = L_1 \cup \cdots \cup L_n$. Thus, the link state $|\cL^n\rangle$ encodes all the coloured link invariants corresponding to the link $\cL^n$ at level $k$. 

We are interested in studying the entanglement structure of these states. To do so, we will
compute the entanglement entropy corresponding to partitioning the $n$-component link into an $m$-component sub-link $L_1 \cup L_2 \cup \cdots \cup L_m$ and its complement $L_{m+1} \cup \cdots \cup L_n$
\beq
S_{EE;\,(L_1, \cdots, L_m| L_{m+1}, \cdots, L_n)} = -\mathrm{Tr}_{L_{m+1}, \cdots , L_n}(\rho\,\mathrm{ln}\,\rho),\;\;\; \rho = \frac{1}{\langle \cL_n |\cL_n\rangle }\mathrm{Tr}_{L_1,\cdots, L_m}|\cL^n\rangle\langle \cL^n| \, ,
\eeq
where by tracing over $L_i$ we mean tracing over the Hilbert space of the torus boundary corresponding to the circle $L_i$. We will  interchangeably use the notation $(L_1,\cdots, L_m | L_{m+1},\cdots, L_n)$ or $(m|n-m)$ to denote such bi-partitions; the former notation makes explicit which components of the link will be traced over.
%which circles inside the link we intend to reduce the state over. 

This computation can be carried out generally in the case of $G=U(1)_k$; we do this Section \ref{sec3}. In the non-Abelian case (we take $G=SU(2)_k$ for simplicity), the general computation is more challenging, and so we will proceed by considering various examples of two- and three-component links in Section \ref{sec4}. This will help us extract useful lessons about the topological entanglement structure of these link states. 

\myfig{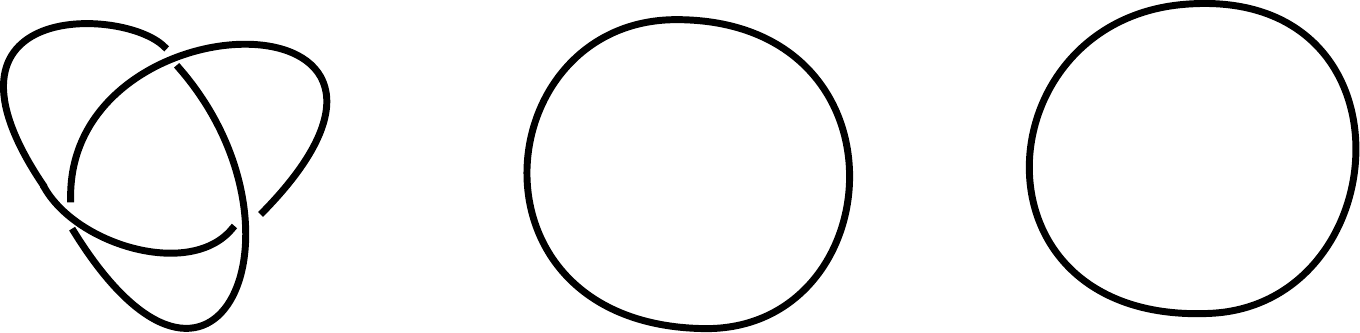}{7}{\small{\textsf{Three unlinked knots.}}}

However, two important facts are immediately obvious:
\begin{itemize}
\item
% There is one somewhat trivial case which we can discuss in full generality, and which clarifies an essential feature of these entanglement entropies. 
Take the link $\cL^n$ to be $n$ un-linked knots (see Figure \ref{fig:fig4.pdf}). In this case, it is well-known that the coloured link-invariant in equation \eqref{cli} factorizes:
\beq
 \frac{C_{\mathrm{unlink}}(j_1, \cdots, j_n)}{C_0} = \prod_{i=1}^n \frac{C_{L_i}(j_{i})}{C_0}
 \eeq
 where $C_0 = \mathcal{S}^0_0$ is the partition function of Chern-Simons theory on $S^3$. It is then clear that the state $|\cL^n\rangle$ is a product state
 \beq
 |\cL^n\rangle \propto |L_1\rangle \otimes |L_2 \rangle \otimes \cdots \otimes |L_n\rangle
 \eeq
and hence the state $|\cL^n\rangle$ is completely unentangled. This is our first hint that the quantum entanglement of link states captures aspects of the topology  of the corresponding links. Specifically, {\it quantum entanglement of a bipartition of  $\cL^n$ into two components
%into two sub-links 
implies topological linking between the two sub-links}.  For $U(1)_k$ Chern-Simons theory we will also prove a converse in the next section (in terms of Gauss linking), but we have not yet arrived at a proof for general non-Abelian theories.
% The converse statement seems harder to prove for a general Chern-Simons theory,  we will be able to make this completely precise in the case of $U(1)_k$ in the next section. 
\item  Above, we ignored the issue of {\it framing} \cite{Witten:1988hf} of the individual circles comprising the link $\cL^n$. Intuitively, if we replace each of the circles in the link with a ribbon, then the relative linking number between the two edges of the ribbon, or \emph{self-linking}, is ambiguous. In general, to fix this ambiguity we must pick a framing for each circle, and consequently the coloured link invariants are really defined for framed links. However a different choice of framing of, let's say, the $i^{th}$ circle $L_i$ by $t$ units is equivalent to performing a $t$-fold Dehn twist on the corresponding torus.  This  corresponds to a local unitary transformation on the corresponding link state:
\beq
|\cL^n\rangle \to \left(1 \otimes 1 \cdots \otimes\mathcal{T}_i^t \otimes 1 \cdots \otimes 1\right) |\cL^n\rangle
\eeq
where $\mathcal{T}_i$ is a Dehn-twist on the $i^{th}$ torus. Local unitary transformations of this type do not affect the entanglement entropies we are interested in. Hence, {\it  the entanglement entropies are framing-independent link invariants.} 
\end{itemize}

\section{The Abelian case: $G=U(1)_k$}\label{sec3}
In this section we will compute the entanglement entropy for arbitrary bi-partitions of a generic $n$-component link in $U(1)_k$ Chern-Simons theory. As warm-up, we will start with two-component links, and then  build up to the general case.  

\subsection{Two-component links}
The main result we will use throughout this section is that if we have an $n$-component link $\cL^n$ with charges $q_1, q_2,\ldots, q_n$ placed on the circles $L_1, L_2,\ldots, L_n$ respectively, then the corresponding coloured link invariant in $U(1)_k$ Chern-Simons theory is given by \cite{Witten:1988hf}
\beq\label{u1}
C_{\cL^n}(q_1,q_2,\ldots,q_n)\equiv \left\langle W_{-q_1}(L_1)\cdots W_{-q_n}(L_n)\right\rangle_{S^3} = \exp\left(\frac{2\pi i }{k} \sum_{i< j} q_iq_j \ell_{ij}\right)
\eeq
where $\ell_{ij}$ is the Gauss linking number between the circles $L_i$ and $L_j$. When $i=j$, this is interpreted as the self-linking or framing of $L_i$. We will pick $\ell_{ii}=0$ by convention, which is reflected in the above summation.  However, as discussed in the previous section, the entanglement entropies we compute are independent of the choice of $\ell_{ii}$. We note from equation \eqref{u1} that the $C_{\cL^n}$ remains unchanged under shifts by multiples of $k$: $\ell_{ij} \to \ell_{ij} + \mathbb{Z}\, k$. We will therefore assume  that the $\ell_{ij}$ are all chosen such that $0 \leq \ell_{ij}< k$, i.e., $\ell_{ij} \in \mathbb{Z}_k$. 

For a two component link $\cL^2$, equation \eqref{u1} then implies that the wavefunction is
\beq
|\cL^2\rangle = \frac{1}{k}\sum_{q_1,q_2} e^{ \frac{2\pi iq_1q_2}{k}\ell_{12}} |q_1\rangle \otimes |q_2\rangle
\eeq
where the sum runs over $0$ to $k-1$, i.e., $\mathbb{Z}_k$, and we have introduced a factor of $k^{-1}$ above to normalize the state. If we now wish to compute the entanglement entropy between 1 and 2, the first step is to trace out one of the links:
\beq\label{emdm}
\rho_1 = \mathrm{Tr}_{L_2} |\cL^2\rangle\langle \cL^2| =\frac{1}{k^2} \sum_{q_1, q'_1, p} |q_1\rangle\langle q'_1| e^{2\pi i\frac{(q_1-q'_1)\ell_{12}}{k}p}
\eeq
The sum over $p$ is easy to perform, and we obtain
\beq
\frac{1}{k}\sum_{p=0}^{k-1}e^{2\pi i\frac{(q_1-q'_1)\ell_{12}}{k}p} =\eta_{q_1,q'_1}(k,\ell_{12}) \equiv  \begin{cases} 1 \;\;\;\cdots\;\;\; \ell_{12}(q_1-q'_1) = 0\, (\mathrm{mod}\,k) \\  0\;\;\;\cdots\;\;\; \ell_{12}(q_1-q'_1) \neq 0\, (\mathrm{mod}\,k)\end{cases}
\eeq
The matrix $\eta_{q_1,q'_1}(k,\ell_{12}) $ can be written in the following tensor-product form
\beq
\eta(k,\ell_{12}) = \left(\begin{matrix} 1 & 1 & \cdots & 1\\ 1& 1 & \cdots & 1\\ \vdots  & \vdots &  & \vdots\\1 & 1 & \cdots & 1 \end{matrix}\right)_{(g,g)} \otimes \left(\begin{matrix} 1 & 0 & \cdots & 0\\ 0& 1 & \cdots & 0\\ \vdots  & \vdots &  & \vdots\\0 & 0 & \cdots & 1 \end{matrix}\right)_{\left(\frac{k}{g},\frac{k}{g}\right)} \label{tp}
\eeq
where $g = \mathrm{gcd}(k,\ell_{12})$ and the subscripts on the matrices indicate their dimensions. The eigenvalues of $\eta$ are therefore $\lambda_1=0$ with degeneracy $\left(k-\frac{k}{\mathrm{gcd}(k,\ell_{12})}\right)$, and $\lambda_2 = \mathrm{gcd}(k,\ell_{12})$ with degeneracy $\frac{k}{\mathrm{gcd}(k,\ell_{12})} $. Computing the entanglement entropy from here, we find
\beq 
S_{EE;\,L_1|L_2}(\cL^2) = \mathrm{ln}\,\left(\frac{k}{\mathrm{gcd} (k,\ell_{12})}\right)
\eeq
Thus the entanglement entropy in this case captures  information about the Gauss linking number $\ell_{12}$ filtered by the level of the Chern-Simons theory, namely $\mathrm{gcd}(k,\ell_{12})$. Note from the above formula that the Hopf link (which has $\ell_{12}=1$) is maximally entangled -- this is in fact generally true even in the non-Abelian case, as we will see later.  {\it Thus, the Hopf link is analogous to a Bell pair in quantum information theory}.

\newcommand{\rn}{\boldsymbol{n}}
For later use, it is useful to derive the above expression from a slightly different point of view, using Renyi entropies. The $\rn$th Renyi entropy is defined as
\beq\label{ren}
S_{\rn}(\cL^2) = \frac{1}{1-\rn} \mathrm{ln}\,\mathrm{Tr}_{L_1} \rho_1^{\rn}
\eeq
where $\rn$ is called the Renyi index and the subscript on the trace indicates that we are tracing over the first Hilbert space.  The entanglement entropy is obtained as the limit $\rn \to 1$. From equation \eqref{emdm}, we obtain
\beq \label{re}
S_{\rn} = \frac{1}{1-\rn} \mathrm{ln}\,\left(\frac{1}{k^{\rn}}\sum_{q_1,\cdots, q_{\rn}} \eta_{q_1,q_2}(k,\ell_{12}) \eta_{q_2,q_3}(k,\ell_{12}) \cdots \eta_{q_{\rn},q_1}(k,\ell_{12})\right)
\eeq
where all the sums are over $\mathbb{Z}_k$. The summand is non-zero only provided we satisfy the following conditions
\beqn
\ell_{12}(q_1 - q_2) &=& 0 \,(\mathrm{mod} \,k)\nonumber\\
\ell_{12}(q_2 - q_3) &=& 0\, (\mathrm{mod} \,k)\nonumber\\
 & &\vdots\\
\ell_{12}(q_{\rn} - q_1) &=& 0\, (\mathrm{mod} \,k),\nonumber
\eeqn
in which case it is equal to one. So the sum in equation \eqref{re} is essentially the number of solutions inside $\mathbb{Z}_k^{\rn}$ to the above equations. Suppose we pick an integer $0 \leq q_1 < k$. Then $q_2$ can take on $\mathrm{gcd}(k,\ell_{12}) $ values such that the first of the above conditions is satisfied. Similarly, $q_3$ can take $\mathrm{gcd}(k,\ell_{12}) $ values such that the second condition is satisfied, and so on. The last condition of course is redundant once we satisfy the first $\rn -1 $ of them. Finally, summing over $q_1$, we obtain
\beq
S_{\rn}(\cL^2) = \frac{1}{1-\rn} \mathrm{ln}\,\left(\frac{\mathrm{gcd}(k,\ell_{12})}{k}\right)^{\rn-1} = \mathrm{ln}\,\left(\frac{k}{\mathrm{gcd}(k,\ell_{12})}\right)
\eeq
So we find that the Renyi entropies $S_{\rn}$ are in fact independent of $\rn$. Thus the $\rn \to 1$ limit is trivial, and is equal to the entanglement entropy $S_{EE;\;L_1|L_2}$ computed previously. We will find that the above Renyi trick easier to work with in the general case. 

\subsection{Three-component links}
Let us now move on to the case of 3-component states. Again, we take a generic 3-component link $\cL^3$ and use the coloured link invariants to write down the corresponding state
\beq
|\cL^3\rangle = \frac{1}{k^{3/2}}\sum_{q_1,q_2,q_3} e^{2\pi i \left(\frac{q_1q_2}{k}\ell_{12}+\frac{q_2q_3}{k}\ell_{23}+\frac{q_3q_1}{k}\ell_{13}\right)} |q_1\rangle \otimes |q_2\rangle \otimes |q_3\rangle \, .
\eeq
Let us consider the entanglement entropy for the bi-partition $(L_1|L_2,L_3)$. We trace out links 2 and 3 to obtain the reduced density matrix  over the first factor:
\beq\label{3q}
\rho_1 = \mathrm{Tr}_{L_2,L_3} |\cL^3\rangle\langle \cL^3| =\frac{1}{k} \sum_{q,q'} |q \rangle\langle q'| \,\eta_{q,q'}(k,\ell_{12})\eta_{q,q'}(k,\ell_{13})
\eeq
where $\eta$ is the matrix in (\ref{tp}).  Repeating the arguments in the two-component case, %we  can again write this reduced density matrix as a tensor product of the form given in equation \eqref{tp}. %. Then using the formula
%\beq \label{sf}
%\mathrm{lcm}\left(\frac{k}{\mathrm{gcd}(k,\ell_{12})},\frac{k}{\mathrm{gcd}(k,\ell_{13})} ,\cdots, \frac{k}{\mathrm{gcd}(k,\ell_{1n})}\right) = \frac{k}{\mathrm{gcd}(k,\ell_{12},\ell_{13},\cdots,\ell_{1n})}
%\eeq
it is easy to show that the non-zero eigenvalue of the reduced density matrix is $\lambda = \frac{\mathrm{gcd} (k,\ell_{12},\ell_{13})}{k}$ with degeneracy $\frac{k}{\mathrm{gcd} (k,\ell_{12},\ell_{13})}$. Thus, the entanglement entropy is given by
\beq \label{ee3}
S_{EE;\,L_1|L_2,L_3}(\cL^3) = \mathrm{ln}\,\left(\frac{k}{\mathrm{gcd} (k,\ell_{12},\ell_{13})}\right)
\eeq

Let us now compute the Renyi entropies for the $(L_1|L_2,L_3)$ partition. From equations \eqref{ren} and \eqref{3q}, we obtain
\beq \label{re2}
S_{\rn}(\cL^3) = \frac{1}{1-\rn} \mathrm{ln}\,\left(\frac{1}{k^{\rn}}\sum_{q_1, \cdots, q_{\rn}} \eta_{q_1,q_2}(k,\ell_{12}) \eta_{q_1, q_2} (k, \ell_{13}) \cdots   \eta_{q_{\rn},q_1}(k,\ell_{12}) \eta_{q_{\rn}, q_1} (k, \ell_{13}) \right)
\eeq
Following arguments similar to the two-component case, the sum only receives contributions from terms which satisfy 
\beqn\label{3c}
\ell_{12}(q_1 - q_2) &=& 0 \,(\mathrm{mod} \,k),\;\;\;\ell_{13}(q_1 - q_2) = 0 \,(\mathrm{mod} \,k)\nonumber\\
\ell_{12}(q_2 - q_3) &=& 0 \,(\mathrm{mod} \,k),\;\;\;\ell_{13}(q_2- q_3) = 0 \,(\mathrm{mod} \,k)\nonumber\\
 & &\vdots \\
\ell_{12}(q_{\rn} - q_1) &=& 0\, (\mathrm{mod} \,k),\;\;\; \ell_{13}(q_{\rn} - q_1) = 0\, (\mathrm{mod} \,k)\nonumber
\eeqn
where we note that the number of constraints has doubled as compared to the two-component case. The sum in equation \eqref{re2} is then precisely equal to the number of integer-valued solutions in $\mathbb{Z}^{\rn}_k$ to the congruences \eqref{3c}. To find these solutions, once again we pick some $0\leq q_1 < k$. Then the number of choices for $q_2$ corresponds to the number of solutions to the equations
\beqn \label{con}
\ell_{12} \,x = 0\,( \mathrm{mod}\, k ),\qquad
\ell_{13} \, x = 0\,( \mathrm{mod}\, k ).
\eeqn
which is $\mathrm{gcd}(k, \ell_{12}, \ell_{13})$. Similarly, $q_3$ can be picked in $\mathrm{gcd}(k, \ell_{12}, \ell_{13})$ ways, and so on. Finally, summing over $q_1$, we obtain
\beq\label{re23}
S_{\rn}(\cL^3) = \mathrm{ln}\,\left(\frac{k}{\mathrm{gcd} (k,\ell_{12},\ell_{13})}\right)
\eeq
which agrees with eq. \eqref{ee3}. Once again, we note that the Renyi entropies are independent of the Renyi index $\rn$. 

It is useful to make the above counting procedure more systematic. Let us define the \emph{linking matrix} for the $(L_1|L_2,L_3)$ partition as (the general definition is given below, eq. \eqref{gl})
\beq
\bs{G} = \left(\begin{matrix} \ell_{12} \\ \ell_{13} \end{matrix} \right)
\eeq
We interpret $\bs{G}$ as a matrix over the field $\mathbb{Z}_k$, i.e., as a map $\bs{G}: \mathbb{Z}_k  \to \mathbb{Z}_k\times \mathbb{Z}_k$. Then, the Renyi entropy, eq. \eqref{re23}, can be rewritten in terms of the  linking matrix as
\beq \label{re24}
S_{\rn} = \mathrm{ln}\,\left(\frac{k}{|\mathrm{ker}\,\bs G |}\right)
\eeq
where by $|\mathrm{ker}\,\bs G |$ we mean the number of solutions in $\mathbb{Z}_k$ to the congruences \eqref{con}, \emph{including} the zero solution. In the present case, clearly $|\mathrm{ker}\,\bs G | = \mathrm{gcd}(k,\ell_{12}, \ell_{13})$.

We can also compute other information theoretic quantities in this setup, for instance the \emph{mutual information} between, say, the links $L_1$ and $L_2$
\beq
I(L_1, L_2) = S_{EE}(L_1) + S_{EE}(L_2) - S_{EE}(L_1\cup L_2) = \mathrm{ln}\left( \frac{\mathrm{gcd}(k,\ell_{13}, \ell_{23})}{\mathrm{gcd}(k,\ell_{12}, \ell_{13}) \mathrm{gcd}(k,\ell_{12}, \ell_{23})} k\right)
\eeq
where $S_{EE}(L_1)\equiv S_{EE; L_1|L_2,L_3}, S_{EE}(L_2)\equiv S_{EE; L_2|L_1,L_3}$, and $S_{EE}(L_1\cup L_2)\equiv S_{EE; L_1,L_2|L_3}$.
A standard result in quantum information theory is that the mutual information is a positive semi-definite quantity. This positivity condition together with equation \eqref{ee3} then translates to the identity
\beq
\frac{\mathrm{gcd}(k,\ell_{12}, \ell_{13}) \mathrm{gcd}(k,\ell_{12}, \ell_{23})}{\mathrm{gcd}(k,\ell_{13}, \ell_{23})}\leq k
\eeq
which is easily verified.
%The Renyi entropy as written in \eqref{re24} generalizes to the case of an arbitrary bi-partition of a generic $n$-component link.

\subsection{$n$-component links}
Let us now consider an $n$-component link $\cL^n$. We wish to compute the entanglement entropy for a $(m|n-m)$ bipartition between the $m$-component sublink consisting of the circles $(L_1,L_2,\cdots L_m)$ and the complement sub-link consisting of $(L_{m+1},\cdots, L_n)$. We may choose  $m\leq n- m$ without loss of generality. Tracing over the links $(L_{m+1},\cdots, L_n)$, we obtain the reduced density matrix:
\beq
\rho_{1,2\cdots,m} = \frac{1}{k^m}\sum_{q_1\cdots, q_m}\sum_{q_1',\cdots, q_m'}\left( \prod_{i=m+1}^n\eta_{q_1\cdots q_m;q_1'\cdots q_m'}(k,\ell_{1,i},\ell_{2,i}\cdots,\ell_{m,i})\right) e^{i\phi}|q_1\cdots q_m\rangle\langle q_1',\cdots q_m'|
\eeq 
where 
\beq
\eta_{q_1,\cdots,q_m;q_1',\cdots q_m'}(k,\ell_{i1},\cdots,\ell_{i,m}) =\frac{1}{k} \sum_{p}e^{\frac{2\pi i}{k}\left((q_1-q_1')\ell_{1,i} +(q_2-q'_2)\ell_{2,i}+\,\cdots\,+ (q_m-q_m')\ell_{m,i}\right)p} \, ,
\eeq
and
\beq
e^{i\phi} = e^{\frac{2\pi i}{k}\sum_{i<j}^m(q_iq_j-q'_iq'_j)\ell_{ij}}
\eeq
is an unimportant phase which can be eliminated by a unitary transformation on $L_1 \cup L_2\cdots \cup L_m$ (such unitaries acting only on one side of the bi-partition do not affect the entanglement entropy). 
Using precisely the same arguments as before, we can compute the Renyi entropy and we find
\beq
S_{\rn}(\cL^n) = \mathrm{ln}\,\left(\frac{k^m}{|\mathrm{ker}\,\bs G|}\right) \, ,
\eeq
where $\bs{G}$ here is the appropriate linking matrix across the $(m|n-m)$-partition,
\beq\label{gl}
\bs{G} = \left(\begin{matrix} \ell_{1,m+1} & \ell_{2,m+1}  & \cdots & \ell_{m,m+1} \\ \ell_{1,m+2} & \ell_{2,m+2} &\cdots & \ell_{m, m+2}\\ \vdots& \vdots  & & \vdots \\ \ell_{1,n} & \ell_{2,n}  & \cdots & \ell_{m,n}\end{matrix}\right)%\cdot \left(\begin{matrix} x_{1} \\ x_2 \\ \vdots \\ x_{M}\end{matrix} \right) 
\eeq
and we recall that $\ell_{i,j}$ is the Gauss linking number between $L_i$ and $L_j$, modulo $k$.  As before, the matrix $\bs G$ is  interpreted as a map $\bs{G}: \mathbb{Z}_k^m \to \mathbb{Z}_k^{n-m}$, and so $|\mathrm{ker}\,\bs G|$ is defined as the number of solutions $\vec{x}\in \mathbb{Z}^m_k$ (once again, including the zero solution) to the system of congruences 
\beq
\bs{G} \cdot \vec{x} = 0\,(\mathrm{mod}\,k),
\eeq
which can equivalently be written in terms of Diophantine equations if we so prefer. Once again the Renyi entropies are $\rn$-independent. So we finally arrive at the entanglement entropy (i.e., the $\rn\to 1$ limit of the Renyi entropy) for a generic $n$-component link bi-partitioned into an $m$-component link and its complement:
\beq\label{eegen}
S_{EE; m|n-m}(\cL^n) = \mathrm{ln}\,\left(\frac{k^m}{|\mathrm{ker}\, \bs G|}\right) \, .
\eeq
When $m=1$, it is easy to show that\footnote{We can use $S_{EE}(A) = S_{EE}(A^c)$ to obtain $|\mathrm{ker}\,\bs G^T | = k^{n-2m}|\mathrm{ker}\,\bs G |$. For $m=1$, this gives a very simple proof that the number of solutions to the congruence $a_1x_1+ \cdots a_{n-1}x_{n-1} = 0\,(\mathrm{mod}\,k)$ is equal to $k^{n-2}\,\mathrm{gcd}(k, a_1, a_2,\cdots, a_{n-1})$, a result found in standard number theory texts \cite{hua}. } 
\beq
|\mathrm{ker}\,\bs G|= \mathrm{gcd}(k, \ell_{12}, \ell_{13}\cdots, \ell_{1n}),
\eeq
and consequently we have a completely explicit formula for the entanglement entropy. For $m>1$, we do not know of such an explicit formula for $|\mathrm{ker}\,\bs G|$. Nevertheless, as a demonstration of the usefulness of equation \eqref{eegen} we can compute an interesting information theoretic quantity called the \emph{tri-partite mutual information}:
\beq
I_3(L_1,L_2, L_3) = I(L_1, L_2) + I(L_1, L_3) - I(L_1, L_2 \cup L_3)
\eeq
in, for instance, a four-component simple chain, for which $\ell_{12} = \ell_{23} = \ell_{34}=1$ while the rest of the linking numbers vanish. A direct computation shows that in this case
\beq
I_{3} = -\mathrm{ln}\,k < 0 
\eeq
thus indicating genuine tri-partite entanglement in this state. However, the mutual information in these link states does not satisfy monogamy, namely it is possible to construct explicit examples where $I_3 >0$. For instance, this is the case if we take $\ell_{i,j}=1$ for all $i\neq j$, in which case one finds $I_3 = \mathrm{ln}\, k$. A more complete investigation of multi-partite entanglement and the entropy cone in this system will be left to future work.

We are now in a position to answer the following question: what type of topology in a link is detected by the Abelian entanglement entropy? It is clear from the definition \eqref{gl}, that if the Gauss linking matrix $\bs{G}$ vanishes (i.e., $\bs{G} = 0\, (\mathrm{mod}\, k)$), then $|\mathrm{ker}\,\bs G| = k^m$. Consequently, the above expression for $S_{EE; m|n-m}$ implies that the entanglement entropy vanishes. Conversely, if the entropy $S_{EE;m|n-m}$ vanishes, then this implies that $|\mathrm{ker}\,\bs G| = k^m$. In other words, every point in $\mathbb{Z}^m_k$ lies in the kernel of $\bs G$. By applying this condition to special points like $(1,0,0,\cdots, 0),\;(0,1,0\cdots,0)$ etc., we then learn that all the elements of $\bs{G}$ are $0\,(\mathrm{mod}\,k)$. Hence, the linking matrix vanishes, modulo $k$. Therefore, we have proven that {\it the quantum entanglement entropy in $U(1)_k$ Chern-Simons theory for an $(m|n-m)$ bi-partition of a generic $n$-component link vanishes if and only if the corresponding linking matrix $\bs G$ vanishes (modulo $k$).} In this sense, the entanglement entropy in $U(1)_k$ Chern-Simons theory detects Gauss linking modulo $k$.

\section{Non-Abelian case: $G = SU(2)_k$}\label{sec4}
In this section, we will compute the multi-boundary entanglement entropies in the case of a non-Abelian group, $SU(2)_k$. In contrast to the $U(1)_k$ case, the calculation of the entropies cannot be carried out in complete generality.  So our strategy will be to  work out the entropies for several interesting cases of two- and three-component links, and will then discuss general lessons from these examples. 

\subsection{Two-component states}
The simplest non-trivial two-component link is the Hopf link (Figure \ref{fig:fig5.pdf}), denoted by $2^2_1$ in Rolfsen notation.
\myfig{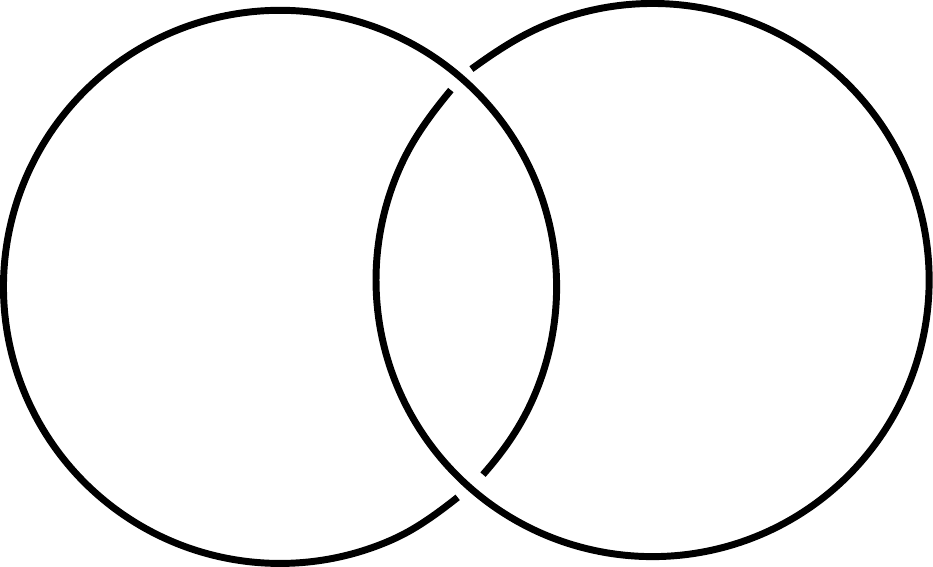}{5}{\small{\textsf{The Hopf-link.}}}
It is possible to evaluate the entanglement entropy in the corresponding state $|2^2_1\rangle$ in several different ways. In fact, the coloured link invariants that define the wavefunction, $C_{2^2_1}(j_1, j_2)$, are given by the modular $\cS$-matrix elements \cite{Witten:1988hf}
\beq
C_{2^2_1}(j_1, j_2) = \cS_{j_1 j_2} \, ,
\eeq
where recall that $\cS$ implements the global diffeomorphism $\tau \to -\frac{1}{\tau}$ on the torus, and for $SU(2)_k$ is explicitly given by
\beq
\cS_{j_1 j_2} = \sqrt{\frac{2}{k+2}} \sin\left(\frac{(2j_1+1)(2j_2+1)\pi}{k+2}\right)
\eeq
The only property of $\cS$ which is relevant presently is that it is unitary. Using this property, it is a simple exercise to show that the normalized reduced density matrix after tracing out the first link is given by
\beq
\rho_2 (2^2_1) = \frac{1}{\langle 2^2_1 | 2^2_1 \rangle} \mathrm{Tr}_{L_1}\,|2^2_1\rangle \langle 2^2_1| = \frac{1}{\mathrm{dim} (\cH(T^2))}\sum_{j} |j\rangle \langle j| \label{hl}
\eeq
Consequently, one finds the entanglement entropy 
\beq
S_{EE}(2^2_1)  = \mathrm{ln}\;\mathrm{dim} (\cH(T^2)) = \mathrm{ln}\,(k+1)
\eeq
which implies that the Hopf link state is maximally entangled. {\it In other words, the Hopf link is analogous to a Bell pair in quantum information theory.} We encountered this fact in the $U(1)_k$ case as well. The same result can also be obtained using the replica trick.   The link complement corresponding to the Hopf link  is $T^2 \times I$, where $I$ is an interval.  Hence, replicating the manifold makes a longer interval, and taking the trace turns the interval into a circle.  Thus, the Renyi entropy essentially amounts to computing the log of the partition function over $S^1\times T^2$; a direct computation then yields the above result. 

%\myfig{fig4.pdf}{5}{\small{\textsf{The reduced density matrix.}}}

%While the above computation was fairly straight-forward, it is useful to take another slightly different route, which might turn out to be somewhat easier to use in the case of some more complicated links. The idea is to directly compute matrix-elements $D_{2^2_1}$ of the required (non-normalized) density matrix 
%\beq
 %\rho^{(0)}_2(2^2_1) =  \mathrm{Tr}_1\,|2^2_1\rangle \langle 2^2_1|= \sum_{j_1, j_1'} D_{2^2_1} (j_1, j'_1) |j_1\rangle \langle j_1'|
%\eeq
%A little bit of mental jugglery shows that the above matrix-element $D_{2^2_1}$ is proportional to the expectation value of two disentangled Wilson loops on $S^1\times S^2$, both running along the non-contractible $S^1$, and with the representation $j_1$ on one of the loops and $j'_1$ on the other (see Figure 5). Therefore, this matrix element is precisely the dimension of the Hilbert space on $S^2$ with one $j_1$ and $j'_1$ insertion, which is $\delta_{j_1, j_1'}$. Normalizing the density matrix, one then recovers the same result as in equation \eqref{hl}. \ack{I think I am being careless about distinguishing between $j$ and its dual $\bar{j}$, or equivalently the relative orientation of loops. }

\myfig{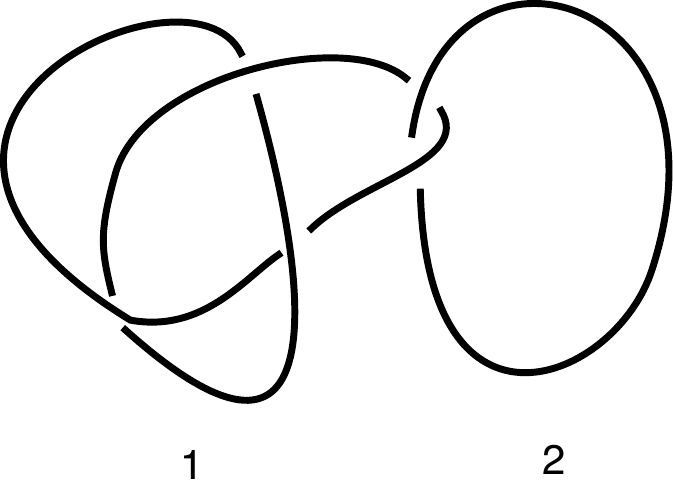}{5}{\small{\textsf{A link between a trefoil knot and an unknot, i.e., the connected sum of the trefoil knot with the Hopf link. }}}

Having studied the Hopf link, it is natural to ask what happens if we replace the individual unknots inside the Hopf link with more complicated knots. In other words, given two knots $K_1$ and $K_2$, what is the link state corresponding to ``Hopf-linking'' these two knots together? (see for instance Figure \ref{fig:fig6.pdf} which illustrates this link for the case of $K_1$ being a trefoil and $K_2$ being an unknot). More precisely, we are asking for the link state corresponding to the \emph{connected sum} $K_1+2^2_1+K_2$ (see \cite{Witten:1988hf} for further details).\footnote{Such a connected sum is not unique in general, but does not apply in the case we're studying.} It is a simple matter (again following \cite{Witten:1988hf}) to write down the state corresponding to this connected sum:
\beq
|K_1+ 2^2_1+ K_2\rangle = \sum_{j_1,j_2} \frac{C_{K_1}(j_1) }{\cS_{0j_1}}\, \cS_{j_1j_2} \,\frac{C_{K_2} (j_2)}{\cS_{0j_2}} |j_1, j_2 \rangle
\eeq
For simplicity, let us pick $K_2$ to be the unknot. The normalized reduced density matrix over the first component then takes the form
\beq
\rho_1(K_1+ 2^2_1+ K_2) = \sum_{j} p_j |j\rangle \langle j|,\;\;\; p_j = \frac{|\frac{C_{K_1}(j)}{\cS_{0j}}|^2}{\sum_{j'} |\frac{C_{K_1}(j')}{\cS_{0j'}}|^2}
\eeq
and therefore the entanglement entropy in this case is given by
\beq
S_{EE}(K_1+ 2^2_1+ K_2) = -\sum_j p_j\, \mathrm{ln}\,p_j.% ,\;\;\;p_i = \frac{|\frac{C_{K_1}(i)}{\cS_{0i}}|^2}{\sum_j |\frac{C_{K_1}(j)}{\cS_{0j}}|^2}
\eeq
Indeed, if we take $K_1$ to be the unknot as well, then we recover the earlier result for the Hopf link. But in general if $K_1$ is some non-trivial knot, then the entropy of entanglement is smaller. This demonstrates that {\it the non-Abelian entanglement entropy detects knotting of the individual components inside a link, something to which the Abelian theory was insensitive.} 

\myfig{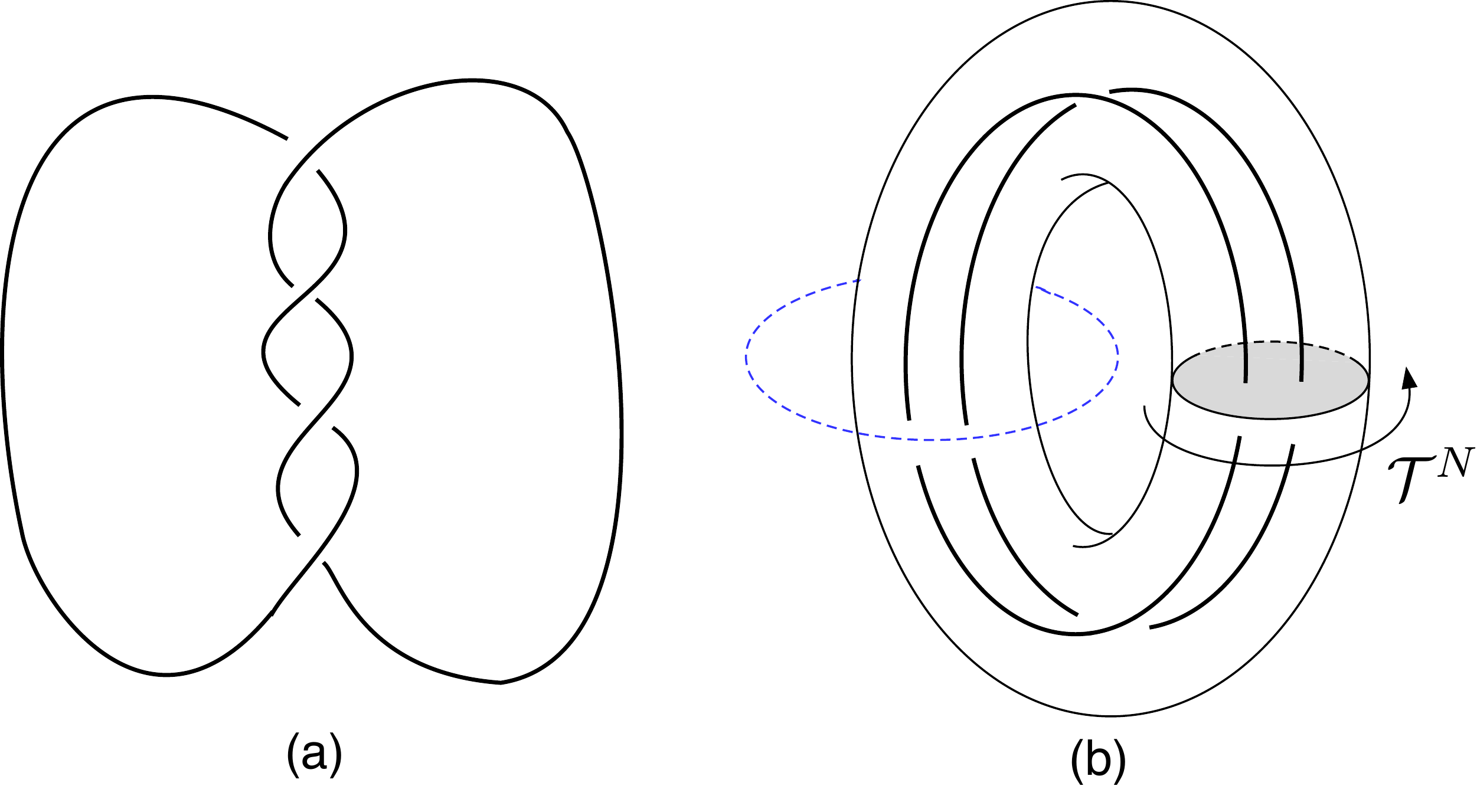}{10}{\small{\textsf{(a) The two component link  $4^2_1$. This is a special case of the family of links $2N^2_1$ with $N=2$. (b) One way to evaluate the corresponding link invariant for general $N$ is to perform surgery along the dashed blue circle. The twisting of the link is accomplished by using a Dehn twist $\cT^N$ as indicated.  }}}

To gain further practice, let us study some additional two-component links. We start with $4^2_1$ (see Figure \ref{fig:fig7.pdf}), which is similar to the Hopf link, but with two twists (or four crossings). In fact, we can instead study the generalization of $4^2_1$ to $2N$ crossings, which we will here denote by $2N^2_1$ (although this is perhaps not the standard terminology). We can explicitly evaluate this state.  To do so, we picture two unlinked circles inside a solid torus and then perform an $N$-fold Dehn-twist on the torus to link the circles together.  Finally, we perform a modular $\cS$ transform and glue the result with an empty solid torus (see Figure \ref{fig:fig7.pdf} (b) for a pictorial explanation of how this is done and \cite{Witten:1988hf} for the details of the general procedure of surgery).  This gives
\beq
|2N^2_1\rangle = \sum_{j_1,j_2}  \sum_m  \left(\cS \cT^N \cS\right)_{0m}\frac{\cS_{j_1m}\cS_{j_2m}}{\cS_{0m}} |j_1, j_2\rangle
\eeq
where we recall that $\cT$ acts by a phase in our basis $\cT|m \rangle = e^{2\pi i h_m }|m\rangle $. The entanglement entropy is therefore given by
\beq
S_{EE} = -\sum_m p_m\, \mathrm{ln}\,p_m,\;\;\; p_m = \frac{\left|\frac{ \left(\cS \cT^N \cS\right)_{0m} }{\cS_{0m}}\right|^2}{\sum_n \left|\frac{ \left(\cS \cT^N \cS\right)_{0n} }{\cS_{0n}}\right|^2} 
\eeq
Since the case $N=1$ (i.e., the Hopf link) is maximally entangled, the entanglement entropy for higher $N$ will generically be smaller (or equal) to the entropy of the Hopf link (see Figure \ref{fig:fig8.pdf}).\footnote{This might seem somewhat counter-intuitive; one might naively have expected that the $N>1$ links are even more entangled. However, it is easy to trace this decrease in entanglement entropy to an increase in the \emph{relative entropy} between the reduced density matrix for $2N^2_1$ and $2^2_1$.  Since the Hopf link was maximally entangled, the only way for this relative entropy to increase is for the $N>1$ links to be less entangled.}

\myfig{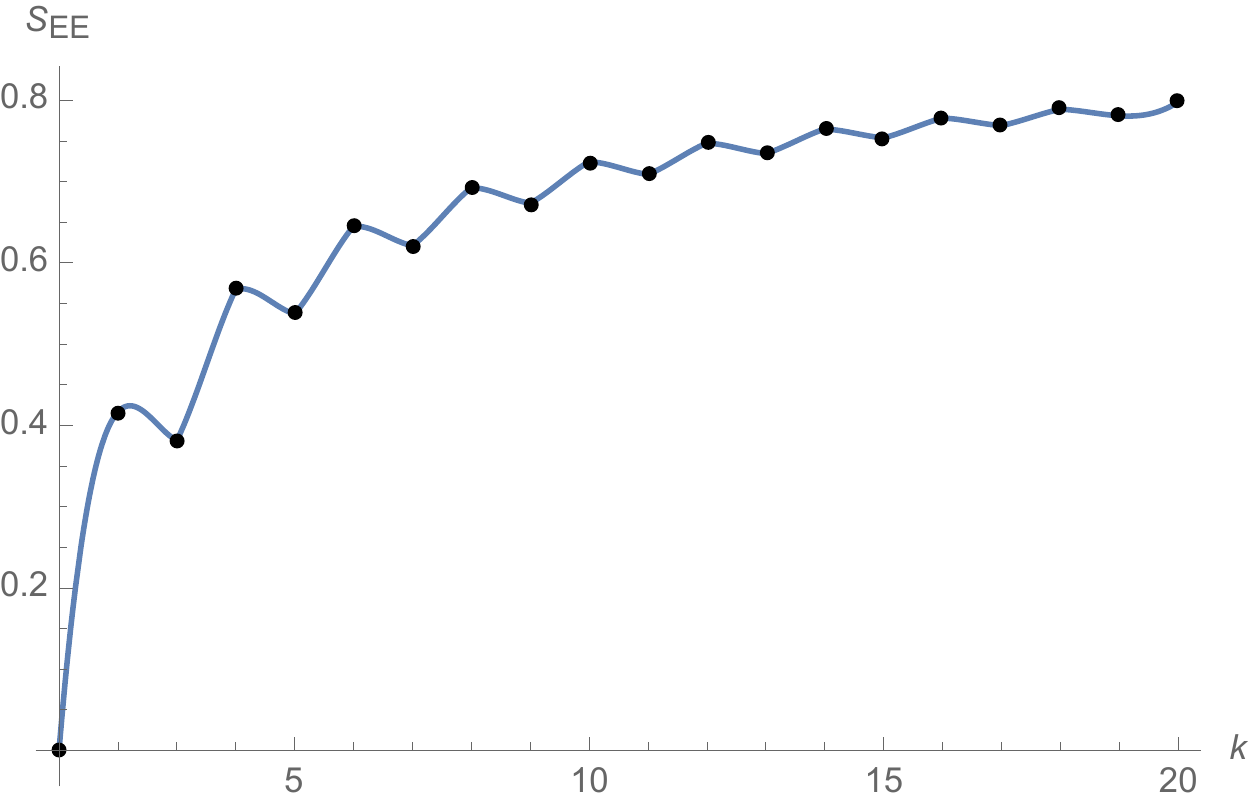}{7}{\small{\textsf{The entanglement entropy of $4^2_1$ as a function of $k$. The blue line is an interpolating curve. }}}

\myfig{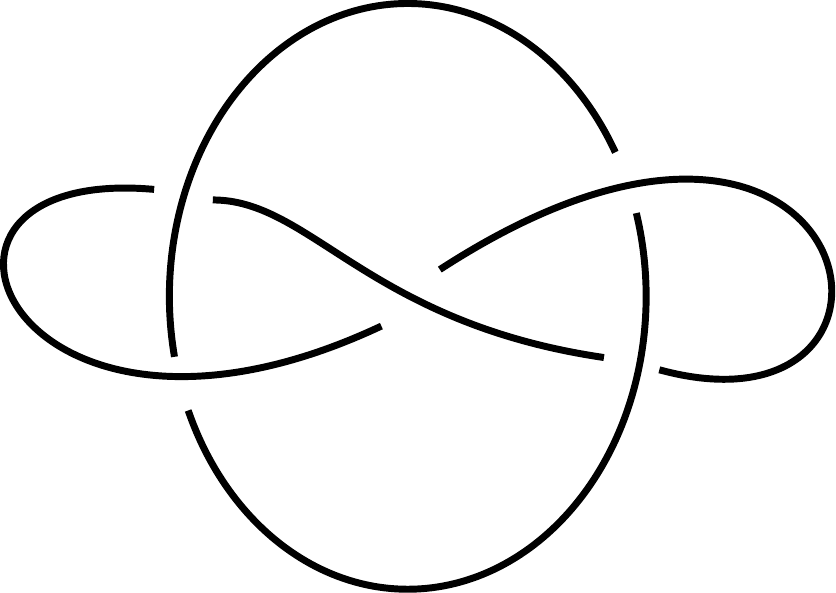}{4.5}{\small{\textsf{The Whitehead link.}}}

Finally, the last two-component link we will study here is $5^2_1$, also called the Whitehead link (Figure \ref{fig:fig9.pdf}).  The Gauss linking number vanishes in this case, but the link is nevertheless topologically non-trivial. The coloured link invariant for the Whitehead link can be computed using a remarkable formula due to K. Habiro \cite{Habiro0, Habiro2008, Gukov:2015gmm}: 
\beq
C_{5^2_1} (j_1, j_2) = \sum_{i=0}^{\mathrm{min}(j_1,j_2)} q^{-\frac{i(i+3)}{4}}(q^{1/2}-q^{-1/2})^{3i}\frac{[2j_1+i+1]!\,[2j_2+i+1]!\, [i]!}{[2j_1-i]!\,[2j_2-i]!\,[2i+1]!}
\eeq
where 
\beq
[x] = \frac{q^{x/2}-q^{-x/2}}{q^{1/2}-q^{-1/2}}, \;\;\; [x]! = [x][x-1]\cdots [1],\;\;\; q= e^{\frac{2\pi i}{k+2}}.\label{nt}
\eeq
The result for the entanglement entropy is shown in Figure \ref{fig:fig10.pdf}. The fact that the Whitehead link has non-trivial entanglement entropy again confirms that {\it the non-Abelian entropy is sensitive not merely to Gauss linking, but to more intricate forms of topological entanglement}. 
\myfig{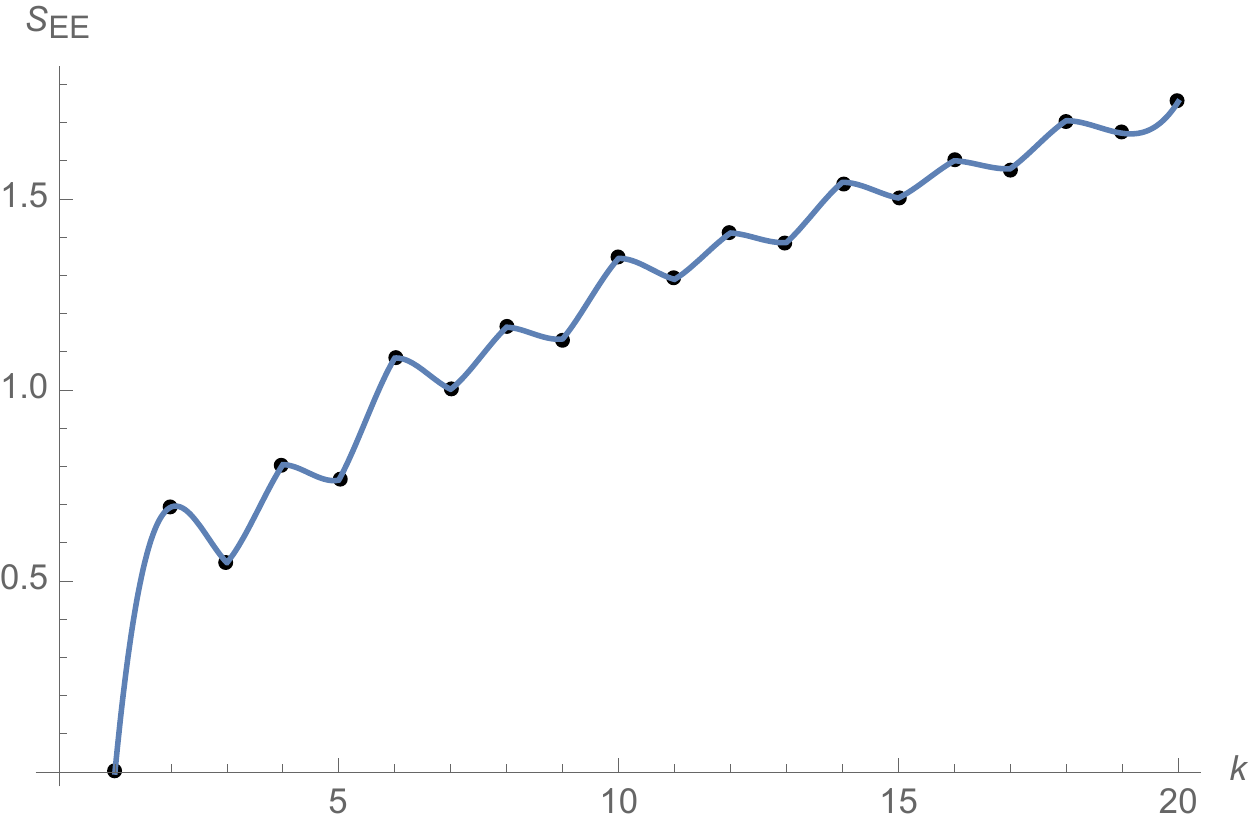}{7}{\small{\textsf{The entanglement entropy for the Whitehead link as a function of $k$. The blue line is an interpolating curve. }}}

There is also a second way to compute the coloured link invariant for the Whitehead link using monodromy properties of conformal blocks of the chiral $SU(2)_k$ WZW model. This method has been explained in detail in \cite{Kaul:1991np, Kaul:1993hb, Kaul:1998ye} and will be reviewed in Appendix A. We merely quote the result here:
\beqn\label{wl}
C_{5^2_1} (j_1, j_2) &= & [2j_1+1]^2[2j_2+1]\sum_{\bs{m, n, p}}  {\lambda}^{-1}_{ p_1,-}(j_1,j_2) \lambda_{p_2,+}(j_1,j_2) {\lambda}^{-1}_{n_1,+}(j_1,j_2){\lambda}^{-1}_{ m_1,-}(j_1,j_2) \lambda_{m_2,+}(j_1,j_2)\nonumber\\
&\times &a_{(\bs 0, \bs p)} \left(\begin{matrix} j_1 &  j_1 \\ j_2 &  j_2 \\  j_1 & j_1 \end{matrix}\right) a_{(\bs n, \bs p)} \left(\begin{matrix} j_1 &  j_2 \\ j_1 &  j_1 \\  j_2 & j_1 \end{matrix}\right)a_{(\bs n, \bs m)} \left(\begin{matrix} j_1 &  j_2 \\ j_1 &  j_1 \\  j_2 & j_1 \end{matrix}\right)a_{(\bs 0, \bs m)} \left(\begin{matrix} j_1 &  j_1 \\ j_2 &  j_2 \\  j_1 & j_1 \end{matrix}\right).
\eeqn
where the $a_{(\bs n, \bs p)}$'s are duality transformations acting on 6-point conformal blocks on $S^2$, and the $\lambda$'s are phases which these blocks pick up under the action of braid generators. In Appendix A all the quantities appearing in equation \eqref{wl} are explained in detail. The relevant point here is that there exists an algorithmic way to compute coloured link invariants using conformal blocks for the Whitehead link, and indeed more generally for arbitrary links. We have also computed the entanglement entropy for the Whitehead link using this second approach for small values of $k$, and we find precise agreement with the results obtained from the Habiro formula.

\subsection{Three-component states}\label{sect3comp}
We now consider a few examples of three-component links and discuss their entanglement structure. Let us begin by considering the link in Figure \ref{fig:fig11.pdf}.  
\myfig{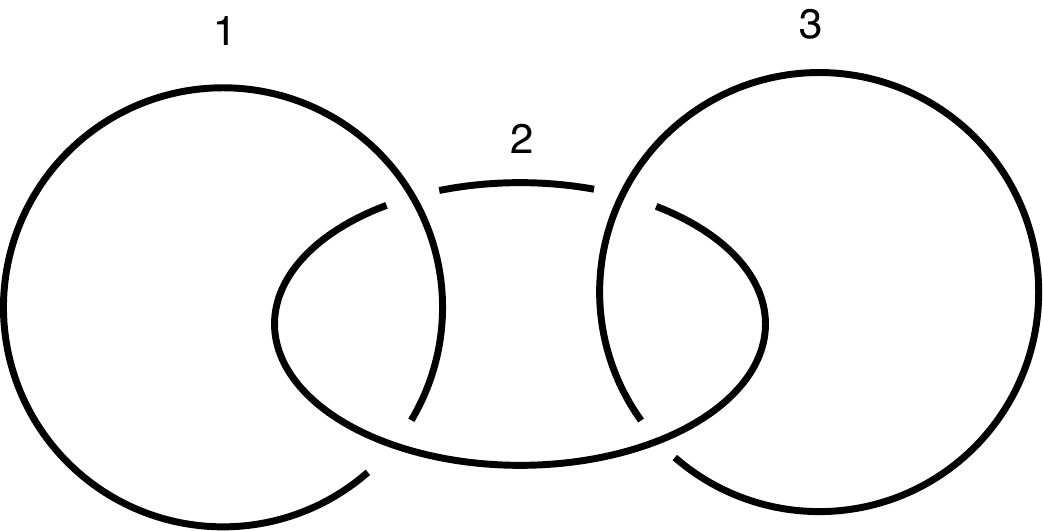}{5.5}{\small{\textsf{A three component link which is the connected sum of two Hopf links. }}}
This link is a connected sum of two Hopf links. Consequently, we can evaluate the link invariant explicitly following \cite{Witten:1988hf}, and we find that the corresponding link state is given by
\beq\label{3cs}
|2^2_1 + 2^2_1\rangle = \sum_{j_1,j_2,j_3,m} \cS_{j_2m}N_{mj_1j_3} |j_1,j_2,j_3\rangle = \sum_{j_1,j_2,j_3} \frac{\cS_{j_1j_2}\cS_{j_3j_2}}{\cS_{0j_2}} |j_1,j_2,j_3\rangle
\eeq
where $N_{ijm}$ is the \emph{fusion coefficient}, namely the dimension of the Hilbert space on $S^2$ with Wilson lines in the representations $i,j,m$ piercing through, or equivalently the number of times the representation $m$ appears in the product of the representations $i$ and $k$.\footnote{Another equivalent way to specify the fusion coefficients is to specify the fusion algebra, which for $SU(2)_k$ is given by:
$$j_1\otimes j_2= |j_1-j_2|, |j_1-j_2|+1, \cdots \mathrm{min}\left(j_1+j_2, k-j_1-j_2\right).$$} We have also used the Verlinde formula \cite{VERLINDE1988360}
\beq
N_{ikm} = \sum_j \frac{\cS_{ij}\cS_{kj}\cS_{mj}}{\cS_{0j}}.
\eeq
So we can compute the entanglement entropies for this state explicitly\footnote{This can be done by changing bases on $L_1$ and $L_3$ to $ |\hat{j}\rangle = \sum_{j'}\cS_{jj'}|j'\rangle$.}, and we find (Figure \ref{fig:fig12.pdf})
\beq\label{sumHopfentspec}
S_{EE; (L_2|L_1,L_3)}(2^2_1+2^2_1) = S_{EE;(L_1|L_2,L_3)}(2^2_1+2^2_1)=-\sum_i p_i\,\mathrm{ln}\,p_i,\;\;\; p_i = \frac{d_i^{-2}}{\sum_j d_j^{-2}}
\eeq
where $d_j = [2j+1]=\frac{\cS_{0j}}{\cS_{00}}$ is the quantum dimension of the representation $j$. %The eigenvalues or Schmidt coefficients $p_i$ are shown in the figure below.
%\myfig{fig12.pdf}{6}{\small{\textsf{The eigenvalues of the reduced density matrix for the connected sum state for $k=35$. }}}
\myfig{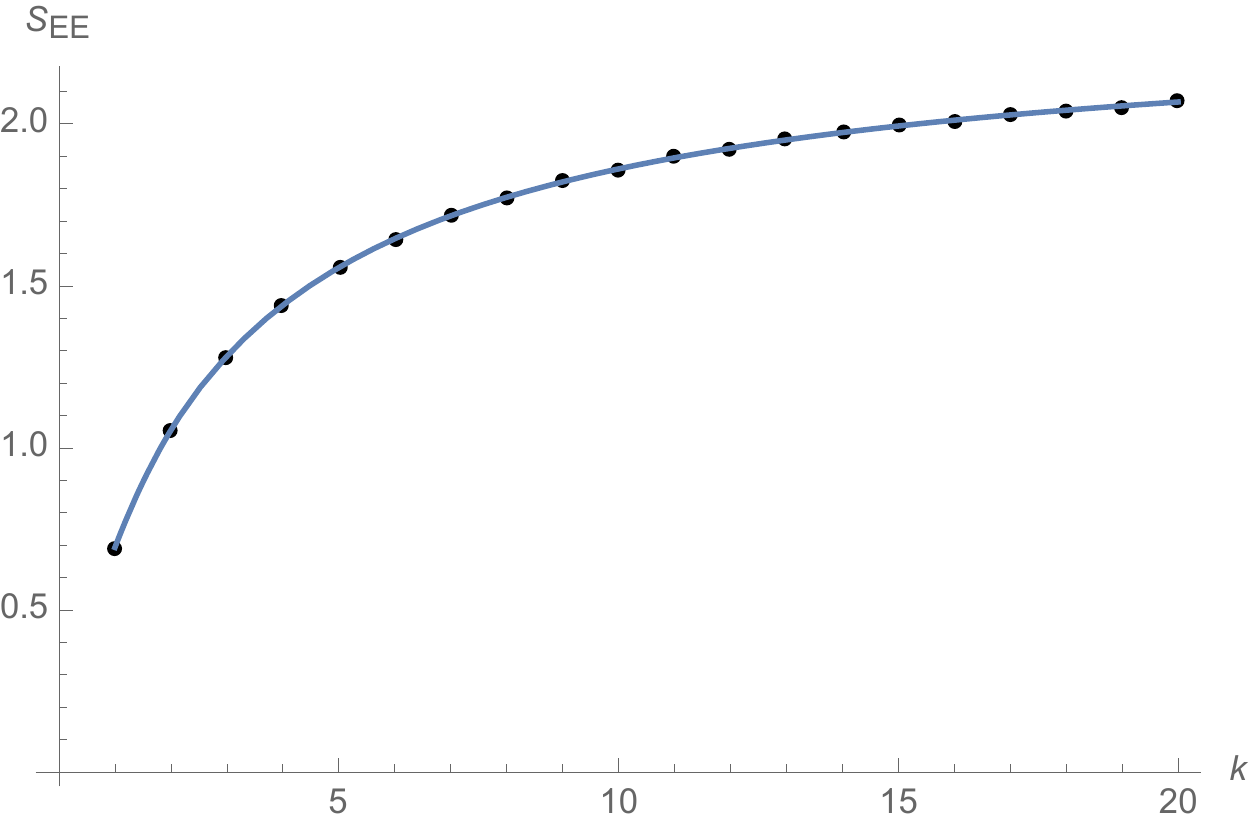}{6}{\small{\textsf{The entanglement entropy $S_{EE; L_2|L_1,L_3}$ for the connected sum of two Hopf links as a function of $k$. }}}
Interestingly, the entropy is independent of which link we trace out. Furthermore, tracing out any of the links leaves us with a \emph{separable} reduced density matrix on the other two links, as can be checked explicitly. In this sense, the above link state has ``GHZ-like'' entanglement. These properties might sound puzzling at first. Indeed, the above discussion makes it clear that the entanglement entropy (and in fact the entanglement spectrum) in this case contains fairly coarse information, and is insufficient to distinguish between the topological linking between for instance the subcomponents 1 and 2 or 1 and 3. Of course, the quantum state has much more fine-grained information which can be potentially extracted by using other probes. For instance, here is one simple-minded way of doing this --- let us define the projector
\newcommand{\bP}{\boldsymbol{P}}
\beq
\bP(L_{\alpha}) = |0\rangle\langle 0 |_{L_{\alpha}}
\eeq
which projects the state on $L_{\alpha}$ to the spin-0 state $|0\rangle$. We can use $\bP(L_{\alpha})$ to further probe the entanglement structure of the state $|2^2_1+2^2_1\rangle$. Acting on various factors of the state \eqref{3cs} with the projector, we get 
\beq
\bP(L_1)|2^2_1 + 2^2_1\rangle = \sum_{j_1,j_2} \cS_{j_1j_2} |0\rangle \otimes |j_1,j_2\rangle 
\eeq
\beq
\bP(L_2)|2^2_1 + 2^2_1\rangle = \sum_{j_1,j_1} \frac{\cS_{j_10}\cS_{j_20}}{\cS_{00}} |j_1\rangle \otimes |0\rangle \otimes |j_2\rangle
\eeq
Note that the latter state is simply a product state. This is easy to understand from the topological structure of the link -- the projector $\bP(L_2)$ essentially erases the second link (that is, a Wilson loop in the spin-0 state is trivial), due to which the link in Figure \ref{fig:fig11.pdf} entirely falls apart into an unlink. So
\beq
S_{EE, L_1|L_3}(\bP(L_2)|2^2_1 + 2^2_1\rangle )=0
\eeq
where we are computing the entanglement entropy of the (pure) state on the links left untouched by the projector. On the other hand, projecting on $L_1$ erases this subcomponent, but the state on the other two links is still non-trivially entangled, mirroring the topological linking in Figure \ref{fig:fig11.pdf}. Indeed, in this case, we find
\beq
S_{EE, L_2|L_3}(\bP(L_1)|2^2_1 + 2^2_1\rangle ) = \mathrm{ln}(k+1)%- \sum_j p_j\,\mathrm{ln}\,p_j,\;\;\; p_j = \frac{\left(\frac{\cS_{mj}}{\cS_{0j}}\right)^2}{\sum_i\left(\frac{\cS_{mi}}{\cS_{0i}}\right)^2} 
\eeq
%We will often refer to this as the \emph{projected} entanglement entropy. 
So the above \emph{projected} entanglement entropies give additional information theoretic measures to probe topological entanglement of links. However, we should emphasize here that we have chosen to project in a particular basis which is natural to the problem; the corresponding entropies are therefore basis-dependent quantities. 
%\ack{Some comments about tracing out v. conditioning...? Jackson, please add your relative entropy calc.} %\footnote{Such a canonical basis only exists for links in $S^3$. However, since any 3-manifold can be obtained by performing Dehn surgery on $S^3$, one should be able to define the projected entropies more generally.}, and the above conclusions are basis-depndent. \ack{It would be nice to have a basis-independent version of the above projection idea.}
\\\\
A basis independent entropic measure that probes how multicomponent links are knotted is the \emph{relative entropy} of the state after being reduced on different links.  Recall that for two states $\rho$ and $\sigma$, the relative entropy is defined by
\begin{equation}
\relent{\rho}{\sigma}=\mTr \left(\rho\ln\rho\right)-\mTr\left(\rho\ln\sigma\right)
\end{equation}
For a three component state $\rho$, computing $\relent{\rho_{L_1}}{\rho_{L_2}}$ gives a basis independent measure of the distinguishability of $\rho$ reduced on link $L_1$ (i.e. where we trace out $L_2$ and $L_3$) against $\rho$ reduced on $L_2$ (i.e. where we trace out $L_1$ and $L_3$).  For instance, considering the chain state (connected sum of Hopf links) $|2^2_1 + 2^2_1\rangle$, the entanglement spectrum of $\rho_{L_1}(2^2_1+2^2_1)$ is the same as $\rho_{L_2}$; however the bases that diagonalize these matrices are different.  Therefore we expect the relative entropy between these two reduced states to be nonzero and indeed we find\footnote{This calculation, along with other various relative entropies can be found in Appendix B.}
\begin{equation}
\relent{\rho_{L_1}(2^2_1+2^2_1)}{\rho_{L_2}(2^2_1+2^2_1)}=\sum_i p_i\left( \mathrm{ln}\,p_i-\sum_j\abs{\cS_{ij}}^2\mathrm{ln}\,p_j\right)
\end{equation}
with $p_j$ being given by \eqref{sumHopfentspec}.  While the projected entropy has  the interpretation of erasing a link, it is not clear that the relative entropy between reduced states has a nice pictorial interpretation.  However, we see that it is a useful entropic measure of the distinguishability of individual components within a given link.

\myfig{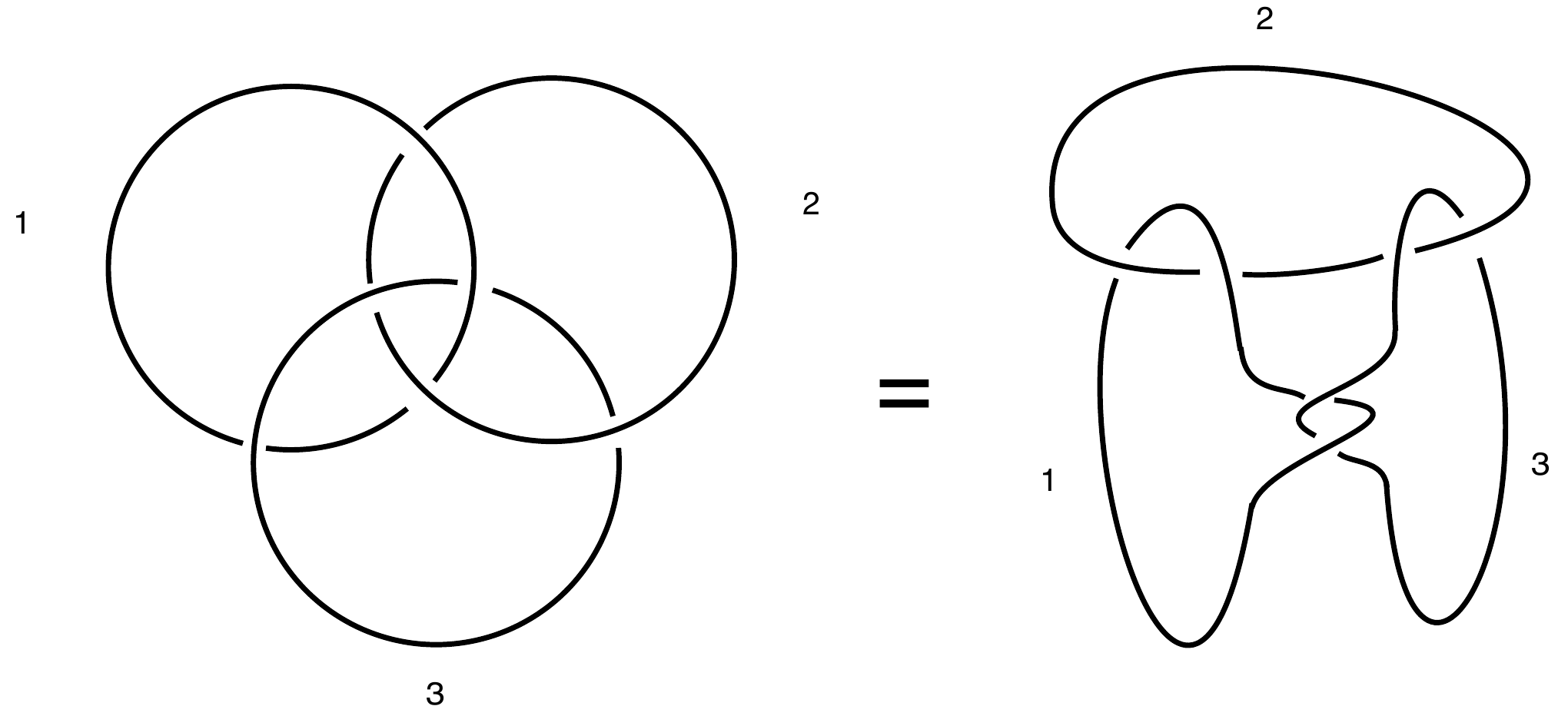}{9}{\small{\textsf{The three component link $6^3_3$. }}}

Let us now consider a slightly more complicated three-component link called $6^3_3$, which is shown in Figure \ref{fig:fig13.pdf}. This differs from the connected sum state we considered previously by a Dehn-twist on a torus surrounding the links 1 and 3. So we can write this  state explicitly as well:
\beqn
|6^3_3\rangle &=& \sum_{j_1,j_2,j_3,m} e^{2\pi i(h_m-h_{j_1}-h_{j_3})}\cS_{mj_2}N_{mj_1j_3} |j_1,j_2,j_3\rangle\\
&=& \sum_{j_1,j_2,j_3}\sum_{m,n} e^{2\pi i(h_m-h_{j_1}-h_{j_3})}\frac{\cS_{mj_2}\cS_{j_1n}\cS_{j_3n}\cS_{mn}}{\cS_{0n}} |j_1,j_2,j_3\rangle \nonumber
\eeqn
where we have used the fact that the Dehn twist acts by a phase in our basis $\cT|m \rangle = e^{2\pi ih_m }|m\rangle $\footnote{We have also corrected for a change in framing that results from the action of $\cT$, although this is not strictly required for our purposes.}. We can simplify the above expressions by using the property $(\cS\cT)^3=1$ (see Section \ref{sec2}), which leads us to
%\beq
%\sum_m e^{2\pi i \left(h_m - \frac{c}{24}\right)} \cS_{im}\cS_{mj} = e^{-2\pi i \left(h_i+h_j - \frac{c}{12}\right)}\cS_{ij}
%\eeq
%\beq
%\sum_m e^{-2\pi i \left(h_m - \frac{c}{24}\right)} \cS_{im}\cS_{mj} = e^{2\pi i \left(h_i+h_j - \frac{c}{12}\right)}\cS_{ij}
%\eeq
%Using these properties, we then find
\beq\label{633}
|6^3_3\rangle = \sum_{j_1,j_2,j_3}\sum_{n} e^{-2\pi i(h_n+h_{j_1}+h_{j_2}+h_{j_3})}\frac{\cS_{j_1n}\cS_{j_2n}\cS_{j_3n}}{\cS_{0n}} |j_1,j_2,j_3\rangle  .
\eeq
Interestingly, the entanglement entropies corresponding to this state are precisely equal to the entanglement entropies for the chain of Hopf links $2^2_1+ 2^2_1$:
\beq
S_{EE;\, L_2|L_1,L_3}(6^3_{3}) = S_{EE;\, L_1|L_2,L_3}(6^3_{3})=S_{EE;\, L_3|L_1,L_2}(6^3_{3})=-\sum_i p_i\,\mathrm{ln}\,p_i,\;\;\; p_i = \frac{d_i^{-2}}{\sum_j d_j^{-2}}
\eeq
Additionally, tracing out any of the links in this state once again leads to a \emph{separable} reduced density matrix on the other two links. This once again implies that this state, like $2^2_1+2^2_1$ has ``GHZ-like'' entanglement (by which we mean that the reduced density matrix obtained by tracing out one of the tori is separable). However, we can distinguish it from the chain of Hopf links state by looking at the projected entropies, namely the entropies after the action of the projector $\bP$. Indeed, it is clear from equation \eqref{633} that all the projected entropies for $6^3_3$ are equal and are given by
\beq
S_{EE, L_2|L_3}(\bP(L_1)|6^3_{3}\rangle ) =S_{EE, L_1|L_3}(\bP(L_2)|6^3_{3}\rangle ) =S_{EE, L_1|L_2}(\bP(L_3)|6^3_{3}\rangle ) = \mathrm{ln}\,(k+1).
\eeq
%where
%\beq
%p_j = \frac{\left(\frac{\cS_{mj}}{\cS_{0j}}\right)^2}{\sum_i\left(\frac{\cS_{mi}}{\cS_{0i}}\right)^2} \nonumber
%\eeq
Notably, the projected entropies for $6^3_3$ are very different from the projected entropies for $2^2_1+2^2_1$, and indeed mirror the topological linking structure of the respective links.  Similarly, a short calculation of the relative entropy between the reduced $6^3_3$ state and the reduced $2^2_1+2^2_1$ state distinguishes these links.  For instance, reducing each link on its second component (i.e. tracing out $L_1$ and $L_3$), we have
\begin{equation}
\relent{\rho_{L_2}(6^3_3)}{\rho_{L_2}(2^2_1+2^2_1)}=\sum_i p_i\left( \mathrm{ln}\,p_i-\sum_j\abs{\cS_{ij}}^2\mathrm{ln}\,p_j\right).
\end{equation}

\myfig{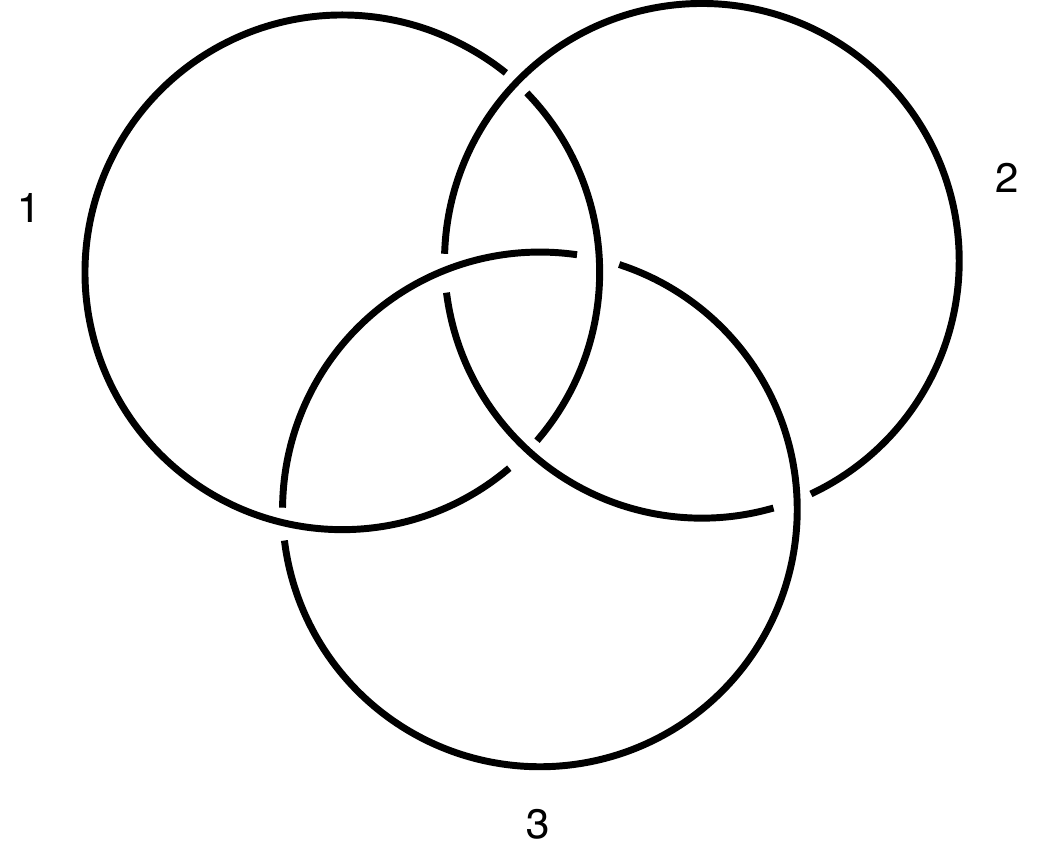}{5}{\small{\textsf{Borromean rings. }}}

Finally, we compute the entanglement entropy for the Borromean rings $6^3_2$ (see Figure \ref{fig:fig14.pdf}). In this case, the coloured link invariants can once again be computed by using Habiro's formula \cite{Habiro0, Habiro2008},\footnote{This formula can be checked explicitly (at least for small values of $k$) using the monodromy of conformal blocks method which is discussed in Appendix A. We find precise agreement in the cases we have checked.} which in this case reads:
\beq
C_{6^3_2}(j_1,j_2,j_3)= \sum_{i=0}^{\mathrm{min}(j_1,j_2,j_3)} (-1)^i (q^{1/2}-q^{-1/2})^{4i}\frac{[2j_1+i+1]!\,[2j_2+i+1]!\, [2j_3+i+1]!\,([i]!)^2}{[2j_1-i]!\,[2j_2-i]!\,[2j_3-i]!\,([2i+1]!)^2}
\eeq
in the notation  introduced in equation \eqref{nt}. Using this formula, it is possible to compute the entanglement entropies for this link as a function of $k$, and the result is shown in Figure \ref{fig:fig15.pdf}.
\myfig{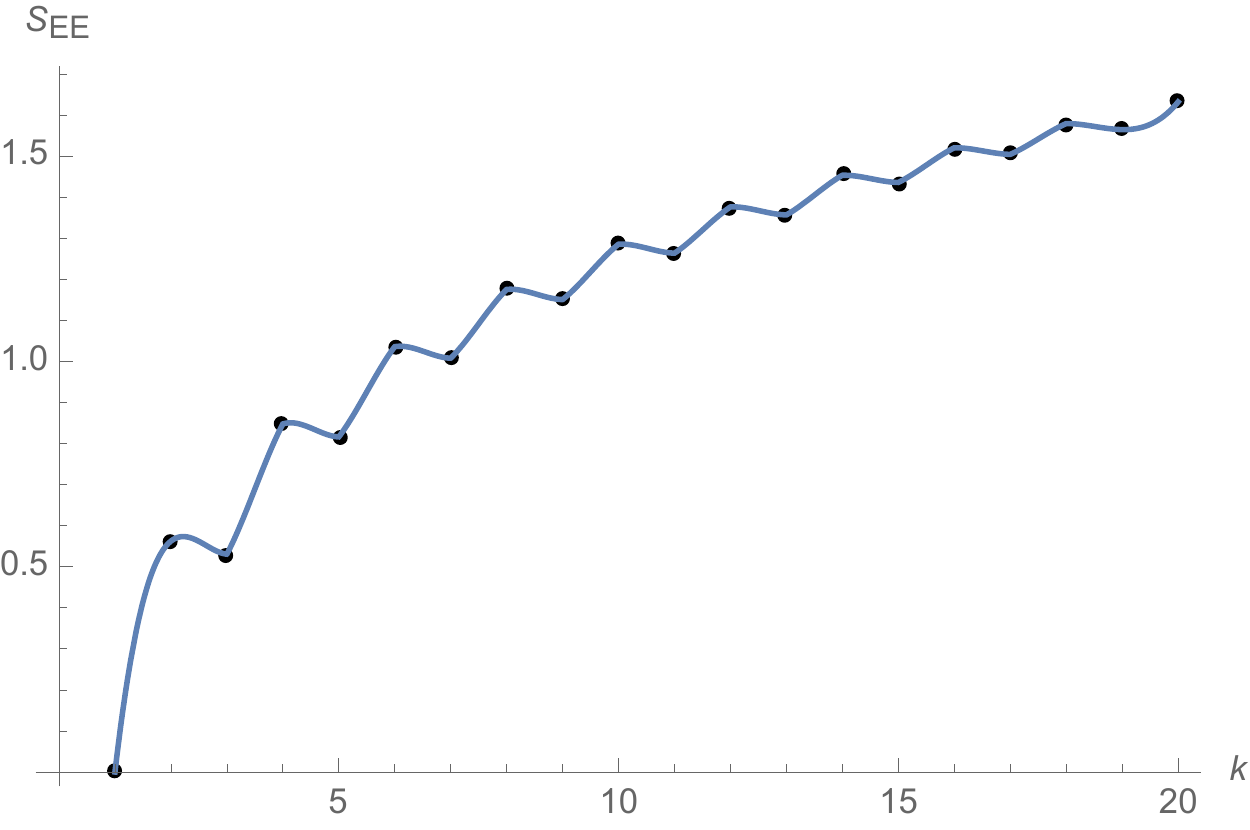}{7}{\small{\textsf{The entanglement entropy for the Borromean rings as a function of $k$ . }}}
Once again, we find that the entropy is non-vanishing in this case. The Borromean rings have trivial Gauss linking between any two circles. Further, they have the special property that if we  erase any circle from the link, the remaining two circles become unlinked; such links are called \emph{Brunnian} links. This latter property can be cast in terms of the projected entropies as the statement that
\beq
S_{EE, L_2|L_3}(\bP(L_1)|6^3_{2}\rangle ) =S_{EE, L_1|L_3}(\bP(L_2)|6^3_{2}\rangle ) =S_{EE, L_1|L_2}(\bP(L_3)|6^3_{2}\rangle ) =0.
\eeq

%from the monodromy properties of conformal blocks on $S^2$. We merely state the result here, and point the reader to Appendix A for further details of this result and the notation:
%\beqn
%C_{6^3_2}(j_1,j_2,j_3) &=& [2j_1+1][2j_2+1][2j_3+1]\sum_{\bs{l},\bs{m},\bs{n},\bs{p},\bs{q}} \lambda_{l_1,-}(j_1,j_2)\lambda_{l_2,-}(j_1,j_3)\lambda_{m_1,-}^{-1}(j_2,j_3)\lambda^{-1}_{n_2,+}(j_1,j_2)\nonumber\\
%&\times & \lambda_{p_0,+}(j_1,j_2)\lambda_{p_1,-}^{-1}(j_1,j_3)\lambda_{q_1,-}^{-1}(j_1,j_2)\lambda_{q_2,-}(j_2,j_3) \nonumber\\
%&\times & a_{(\bs 0, \bs l)} \left(\begin{matrix} j_2 &  j_2 \\ j_1 &  j_1 \\  j_3 & j_3 \end{matrix}\right)a_{(\bs m, \bs l)} \left(\begin{matrix} j_2 &  j_1 \\ j_2 &  j_3 \\  j_1 & j_3 \end{matrix}\right)a_{(\bs m, \bs n)} \left(\begin{matrix} j_2 &  j_1 \\ j_3 &  j_2 \\  j_1 & j_3 \end{matrix}\right)\nonumber\\
%&\times & a_{(\bs p, \bs n)} \left(\begin{matrix} j_2 &  j_1 \\ j_3 &  j_1 \\  j_2 & j_3 \end{matrix}\right)a_{(\bs p, \bs q)} \left(\begin{matrix} j_1 &  j_2 \\ j_1 &  j_3 \\  j_2 & j_3 \end{matrix}\right) a_{(\bs 0, \bs q)} \left(\begin{matrix} j_1 & j_1 \\ j_2 & j_2 \\ j_3 & j_3 \end{matrix}\right)
%\eeqn

Finally, the reduced density matrix for the Borromean rings upon tracing out one of the links (say $L_3$) is \emph{not} separable. The easiest way to see this in the present case is to compute the \emph{entanglement negativity} \cite{PhysRevLett.77.1413, PhysRevA.65.032314} (see also \cite{Rangamani:2015qwa}), which is defined as follows. For a given (possibly mixed) density matrix $\rho$ on a bi-partite system (in the present case on $L_1\cup L_2$), let us start by defining the partial transpose $\rho^{\Gamma}$:
\beq
\langle j_1, j_2 |\, \rho^{\Gamma}\, | \tilde{j}_1, \tilde{j}_2\rangle = \langle j_1, \tilde{j}_2 |\,\rho\, | \tilde{j}_1, j_2\rangle .
\eeq
Then, the number of negative eigenvalues of $\rho^{\Gamma}$ is known to be a good measure of quantum entanglement. A good quantitative way to capture this is the entanglement negativity, which is defined as\footnote{The trace norm is defined as $|| O ||  = \mathrm{Tr}\left(\sqrt{O^{\dagger} O }\right)$.}
\beq
\mathcal{N} = \frac{||\rho^{\Gamma}|| - 1}{2}.
\eeq
More importantly for us, a non-zero value of $\mathcal{N}$ (i.e., $\mathcal{N} > 0 $) necessarily implies that the reduced density matrix is not separable. The negativity for the reduced density matrix on $L_1\cup L_2$ for the Borromean rings is shown in Figure \ref{fig:fig16.pdf}. We find that $\mathcal{N} > 0$ for $k >1$, thus showing that the Borromean rings have a more robust, ``W-like'' entanglement structure (by which we mean that the reduced density matrix obtained by tracing out one of the tori is not separable).

\myfig{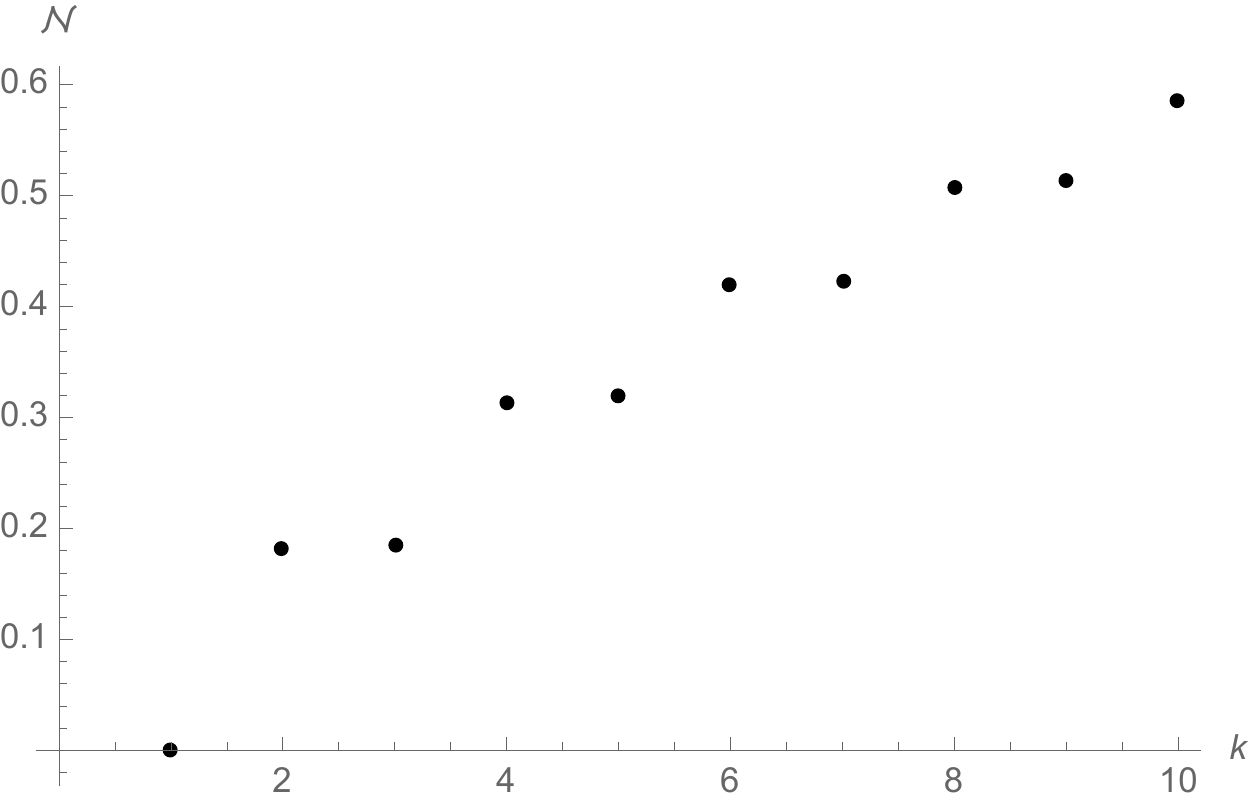}{7}{\small{\textsf{The entanglement negativity between links $L_1$ and $L_2$ upon tracing out $L_3$ for the Borromean rings as a function of $k$ . }}}

\section{Discussion}\label{sec5}
To conclude, we have studied multi-boundary entanglement in Chern-Simons theory for states defined on $n$ copies of a torus $T^2$. We have focussed on the specific class of states prepared by performing the path-integral of Chern-Simons theory on link complements of $n$-component links in $S^3$. For $U(1)_k$ Chern-Simons theory, we gave a general formula for the entanglement entropy of a generic bi-partition of the link into two sub-links. This formula involves the number of solutions of certain congruences (or equivalently Diophantine equations) with coefficients closely related to the Gauss-linking numbers between the two sub-links, and as such relates simple but interesting concepts from quantum information theory, knot theory and number theory. In the non-Abelian $SU(2)_k$ case, we studied the entanglement structure of several two- and three-component links. In particular, we showed that the Hopf link is maximally entangled and thus analogous to a Bell-pair from quantum information theory. We found examples of three component links -- such as $6^3_3$ -- with ``GHZ-like'' entanglement (namely that they have non-trivial, but not necessarily maximal\footnote{Although note that in $U(1)_k$, the $6^3_3$ link additionally also has maximal entanglement under bi-partitions.} entanglement entropies under bi-partitions, but they reduce to separable states upon tracing out one of the links). Finally, we showed that the Borromean rings have a more robust ``W-like'' entanglement structure, namely that they have non-trivial (again, not necessarily maximal) entanglement under bi-partitions, and in addition the reduced density matrix upon tracing out one of the links is not separable. We end with some open questions. 

Generally speaking, a main message of this paper is that quantum information theoretic ideas applied to multi-boundary states in Chern-Simons theory can provide interesting, and potentially powerful tools in the study of knot theory. In this direction, we studied only simple quantities such as entanglement entropies, Renyi entropies, etc., which  turn out to be sums over quantities involving the coloured link invariants. Said another way, the entanglement entropies extract certain coarse-grained framing independent information from the coloured link invariants. In the $U(1)_k$ theory, we showed that these entropies are powerful enough to detect Gauss linking (mod k), namely that the entanglement entropy for a bi-partition vanishes if and only if the Gauss linking matrix between the two sub-links vanishes (mod $k$). In the non-Abelian case, the corresponding statement remains unclear --- it is clear that quantum entanglement implies topological linking, but the converse remains to be shown. In other words, does there exist a link where the coloured link invariants all factorize along a bi-partition, despite non-trivial topological linking between the corresponding sublinks? This is of course also related to a famous question --- do any coloured link invariants detect the unlink? In this context, there are known examples of non-trivial links which the Jones polynomial does not distinguish from the unlink \cite{Eliahou2003155}. It will be interesting to compute the entanglement entropies in these examples. Additionally, it will be of interest to generalize these results to other gauge groups, such as $SU(N)$. 

The discussion above mostly focussed on using quantum information theory to study links. In the opposite direction, we can ask whether knot theory can shed light on unsolved problems in quantum information theory. It is an old idea that quantum entanglement might be interpreted in terms of topological entanglement in links (see for instance \cite{Aravind1997, 1367-2630-4-1-373, Sugita, Solomon} and references therein). We have argued in this paper that multi-boundary states in Chern-Simons theory provide the right framework for realizing this idea. It would be interesting to study whether this connection between quantum entanglement and topological linking can be used effectively in better understanding multi-partite entanglement structures. A first exercise in this direction would be to characterize the entropy cone for multi-boundary states, perhaps in the simpler set-up of $U(1)_k$ Chern Simons theory. It would also be very useful to study the entanglement structure of four and higher component links in the non-Abelian case. 

Finally, it would be interesting to study multi-boundary entanglement in $SL(2,\mathbb{C})$ Chern-Simons theory, which is closely related with quantum gravity in three dimensions. One might expect the multi-boundary entanglement entropy in this context to admit a {\it geometric} description, beyond topology. In fact, it is known that many links (and knots) admit a geodesically complete hyperbolic metric on their link-complements -- such links are called hyperbolic links. For such links, it is conjectured that the logarithm of the reduced $SU(2)$ coloured link invariant with each component carrying the $N$ dimensional representation, evaluated at $q=e^{2\pi i/N}$, asymptotes in the $N\to \infty$ limit to the volume of the hyperbolic metric, a statement which is called the \emph{volume conjecture} \cite{Kashaev:1996kc, Gukov:2003na, Dimofte:2010ep}. Along similar lines, it would be interesting to explore whether the entropies we have defined and computed in this paper also admit a geometric description in terms of the hyperbolic metric on the link complement. Indeed, it would not be unreasonable to hope that the entropy corresponds to the area of some minimal surface (or a horizon in the Lorentzian continuation) in the $k \to \infty$ limit. Of course, this remark is motivated by the Bekenstein-Hawking formula for black-hole entropy, and the Ryu-Takayanagi formula for entanglement entropy in the AdS/CFT correspondence. 

{\bf Note Added}: After this work was completed, we were made aware of the recent work of Salton, Swingle and Walter \cite{Salton:2016qpp}, which has some overlap with our work.  These authors investigate how different states can be prepared on a union of tori in Chern-Simons theory by considering different 3-manifolds with the same boundary.   Their main result is that the states constructed this way in $U(1)_k$ Chern-Simons theory can be interpreted as stabilizer states; this is consistent with the fact that the Abelian Renyi entropies computed in this paper are all equal. They also show that any state in $SO(3)$ Chern-Simons theory can be approximated arbitrarily well through a Euclidean path integral.

\section{Acknowledgments}
We gratefully acknowledge helpful conversations with Matt DeCross, Tom Faulkner, Arjun Kar, Valentijn Karemaker, Sheldon Katz, Romesh Kaul, Christopher Leininger, Aitor Lewkowyzc, Alex Maloney, Henry Maxfield, Djordje Minic, Charles Rabideau, Srinidhi Ramamurthy, Grant Salton, Gabor Sarosi, Brian Swingle, Apoorv Tiwari, Mark Van Raamsdonk and Beni Yoshida. Research supported in part by the U.S. Department of Energy under contracts DE-SC0015655 and  DE-FG02-05ER-41367, and by the Simons Foundation (\#385592, Vijay Balasubramanian) through the It From Qubit Simons Collaboration.  V.B.\ and  O.P.\ also thank the Perimeter Institute for hospitality during the It From Qubit workshop and school.

\appendix

\section{Link invariants from monodromies of conformal blocks}
In this appendix, we review the calculation of coloured link invariants from the monodromy properties of conformal blocks of the $SU(2)_k$ chiral WZW model. We will only review here the recipe for these computations, following \cite{Kaul:1991np, Kaul:1993hb, Kaul:1998ye} (see \cite{Moore:1988qv} for requisite background material); we refer the reader to these papers for further details. Since these techniques are required in this paper for the two special cases of the Whitehead link and the Borromean rings, we will present our discussion in the context of these examples, but the techniques straightforwardly generalize to other links. 

\subsection{Whitehead link}
Our basic ingredients in constructing link invariants will be $S^2$ conformal blocks of chiral vertex operators in $SU(2)_k$ WZW theory. For the case of the Whitehead link (and also Borromean rings), we need the six-point blocks $\phi_{\bs p}$ and $\phi'_{\bs q}$ shown in Figure \ref{fig:fig17.pdf} below. The two different fusion channels correspond to two different choices of a basis for  the Hilbert space of Chern-Simons theory with six Wilson lines piercing through the 2-sphere. In fact, both $\phi_{\bs p}$ and $\phi'_{\bs q}$ are orthonormal bases for the space of six-point conformal blocks on $S^2$ (see Figure \ref{fig:fig17.pdf}), and as such are related by a duality transformation $a_{(\bs p, \bs q)}$:
\beq
|\phi_{\bs p}(j_1,j_2,\cdots, j_6) \rangle = \sum_{\bs q} a_{(\bs p, \bs q)} \left(\begin{matrix} j_1 & j_2 \\ j_3 & j_4 \\ j_5 & j_6 \end{matrix}\right) |\phi'_{\bs q}(j_1,j_2,\cdots, j_6) \rangle
\eeq
\myfig{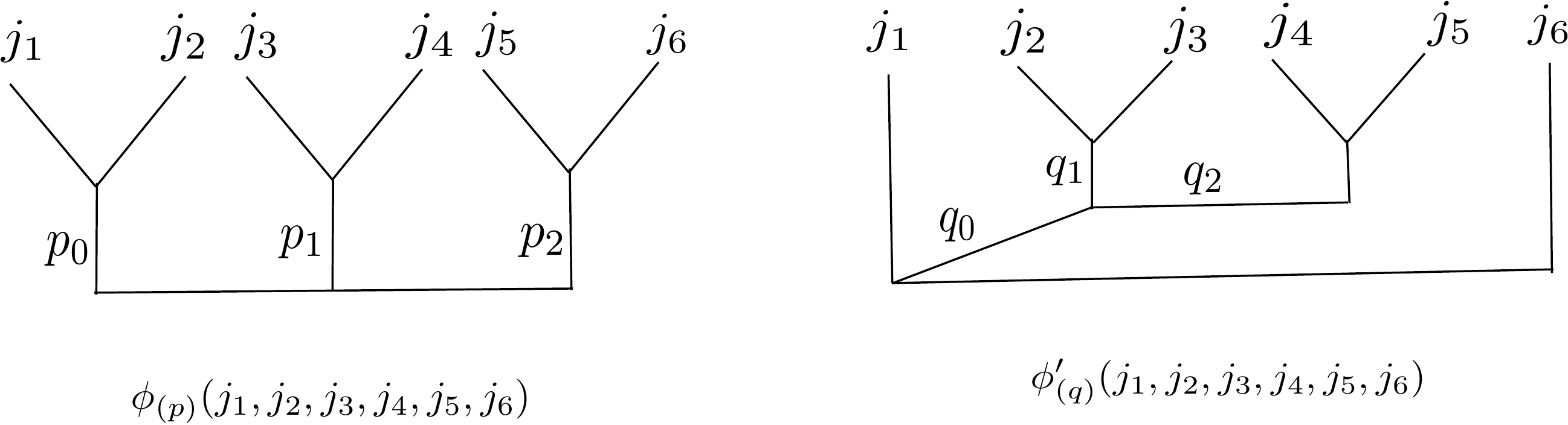}{12}{\small{\textsf{Two different basis for 6-point conformal blocks.}}}

The $a_{(\bs p, \bs q)}$ can also be written in terms of a sequence of four-point duality transformations:
\beq
a_{(\bs p, \bs q)}\left(\begin{matrix} j_1 & j_2 \\ j_3 & j_4 \\ j_5 & j_5\end{matrix}\right) = \sum_t a_{t,p_1}\left(\begin{matrix} p_0 & j_3 \\ j_4 & p_2\end{matrix}\right)a_{p_0,q_1}\left(\begin{matrix} j_1 & j_2 \\ j_3 & t\end{matrix}\right) a_{p_2,q_2}\left(\begin{matrix} t & j_4 \\ j_5 & j_6\end{matrix}\right)a_{t,q_0}\left(\begin{matrix} j_1 & q_1 \\ q_2 & j_6\end{matrix}\right)
\eeq
where $a_{j,l}$ are the fusion matrices for four-point block and are given explicitly by:
\beqn
a_{j,l}\left(\begin{matrix} j_1 & j_2 \\ j_3 & j_4\end{matrix}\right) &=& (-1)^{j_1+j_2-j_3-j_4-2j} \sqrt{[2j+1][2l+1]}\Delta(j_1,j_2,j) \Delta(j_3,j_4,j) \Delta(j_1,j_4,l)\Delta(j_2,j_3,l) \nonumber\\
&\times & \sum_{m\geq 0} (-1)^m [m+1]! \Big\{[m-j_1-j_2-j]![m-j_3-j_4-j]!\nonumber\\
&\times &[m-j_1-j_4-l]![m-j_2-j_3-l]![j_1+j_2+j_3+j_4-m]!\nonumber\\
&\times & [j_1+j_3+j+l-m]![j_2+j_4+j+l-m]!\Big\}^{-1}
\eeqn
where
\beq
\Delta(a,b,c) = \sqrt{\frac{[-a+b+c]! [-b+c+a]! [ -c+a + b]!}{[a+b+c+1]!}}
\eeq
and we have used the notation
\beq
[x] = \frac{q^{x/2}-q^{-x/2}}{q^{1/2}-q^{-1/2}},\;\;\; q = e^{\frac{2 \pi i }{k+2}}
\eeq
\beq
[x]! = [x][x-1][x-2]\cdots [1],\;\;\;[0]! = 1
\eeq

Now coming to the Whitehead link, a plait representation of the link is shown in Figure \ref{fig:fig18.pdf}. 
\myfig{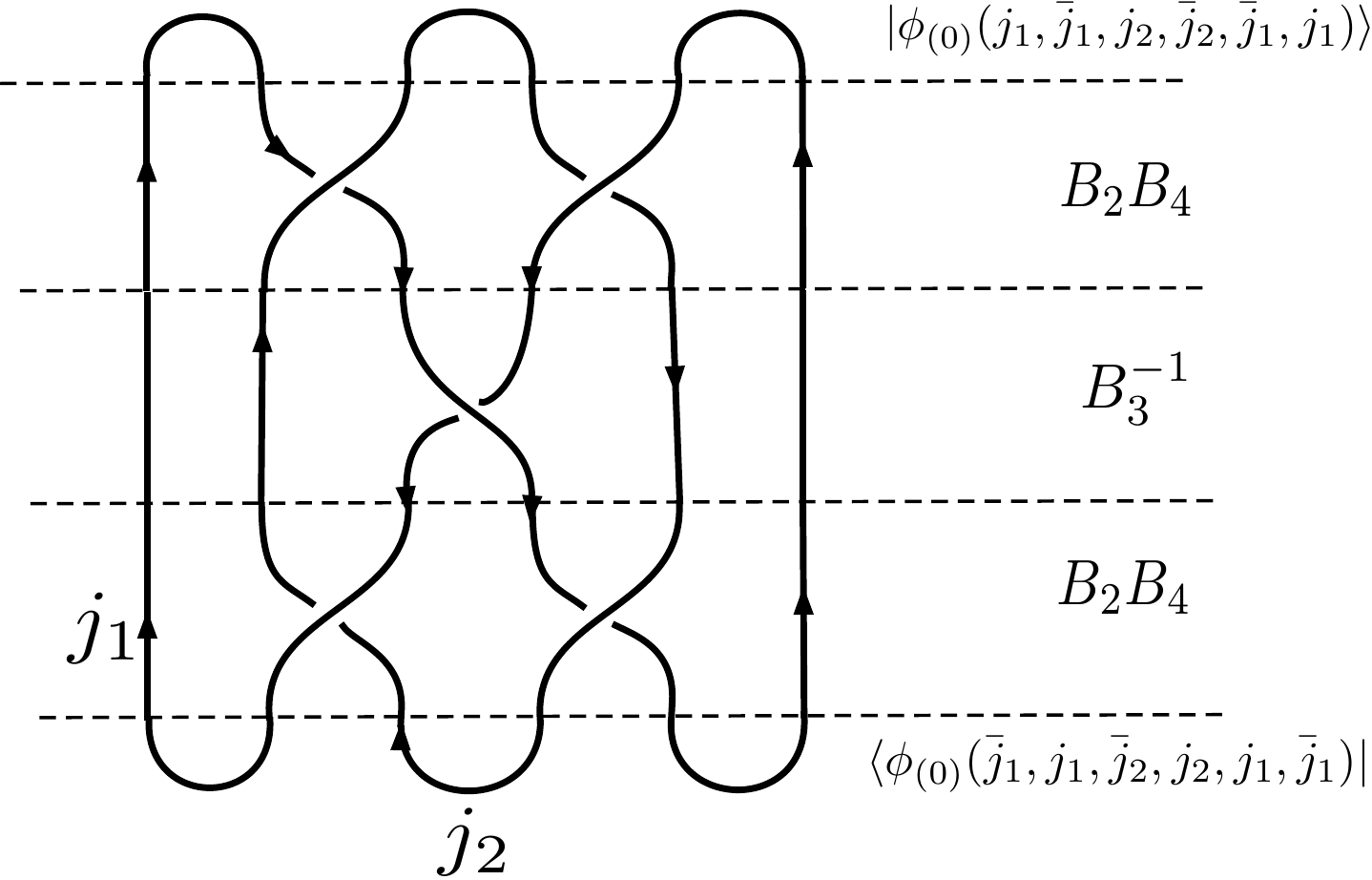}{8}{\small{\textsf{A plait representation of the Whitehead link $5^2_1$.}}}
In order to evaluate this link invariant, we imagine the plait representation as giving a transition amplitude between two states on $S^2$ with six operator insertions. As was argued in \cite{Kaul:1993hb}, the initial state (where by convention we take ``time'' to run from top to bottom) corresponds to the conformal block $\phi_{(0,0,0)}(j_1,\bar{j}_1, j_2, \bar{j}_2, \bar{j}_1, j_1)$, or more precisely
\beq
|\psi_{i}\rangle = [2j_1+1]\sqrt{[2j_2+1]} \left|\phi_{(0,0,0)}(j_1,\bar{j}_1, j_2, \bar{j}_2, \bar{j}_1, j_1)\right\rangle
\eeq
while the final state similarly corresponds to the block $\phi_{(0,0,0)}(\bar{j}_1,j_1, \bar{j_2}, j_2, j_1, \bar{j_1})$ 
\beq
|\psi_{f}\rangle = [2j_1+1]\sqrt{[2j_2+1]} \left|\phi_{(0,0,0)}(\bar{j}_1,j_1, \bar{j_2}, j_2, j_1, \bar{j_1})\right\rangle
\eeq
The operator insertions between the initial and final states implement the braiding of the various strands of the link. The operator $B_{2m+1}$ generates a right handed braid between strand $2m+1$ and $2m+2$, while the operator $B_{2m}$ generates a right-handed braid between the strand $2m$ and $2m+1$. So we can write the Whitehead link invariant as
\beq
C_{5^2_1} (j_1, j_2) =  \langle \psi_{f} | B_2 B_4 B_3^{-1} B_2 B_4 | \psi_{i}\rangle
\eeq
In order to evaluate this amplitude, we need to use the fact that the blocks $|\phi_{(p_0,p_1,p_2)}(j_1,j_2,\cdots, j_6) \rangle$ are eigenstates of odd numbered braiding operators
\beq
B_{2m+1} |\phi_{\bs p}(j_1,j_2,\cdots, j_6) \rangle = \lambda^{\pm 1}_{p_{m},\pm}(j_{2m+1}, j_{2m+2}) |\phi_{\bs p}(j_1,j_2,\cdots, j_6) \rangle
\eeq
where $\bs p = (p_0, p_1,p_2)$, and $\pm$ stands for the relative orientation between the two strands which are bring braided. The other set of blocks $\phi'_{\bs q}(j_1,j_2,\cdots, j_6)$ on the other hand are eigenstates of the even braiding operators
\beq
B_{2m} |\phi'_{\bs q}(j_1,j_2,\cdots, j_6) \rangle = \lambda^{\pm 1}_{q_{m},\pm}(j_{2m+1}, j_{2m+2}) |\phi'_{\bs q}(j_1,j_2,\cdots, j_6) \rangle
\eeq
The eigenvalues appearing above are precisely the monodromies of these conformal blocks, which are given by
\beq
\lambda_{t,\pm} (j_1,j_2)= (-1)^{j_1+j_2-t} q^{\pm \frac{C_{j_1} + C_{j_2} -C_{t}}{2}}
\eeq
where $C_j = j(j+1)$, and the factor $(-1)^{j_1+j_2-t}$ is a symmetry factor.\footnote{Note that \cite{Kaul:1993hb} use eigenvalues which differ from ours by a phase factor. This factor is appended in their case to correct for the change in framing of the link arising from the braiding. But since we are interested in computing entanglement entropies, which as discussed previously are framing independent, we do not need to worry about these framing factors.} As a quick check on this formalism, we can compute the coloured link invariant corresponding to the Hopf link using this method, and we find
\beqn
\frac{\mathcal{S}_{ij}}{\mathcal{S}_{00}} &=& \sum_{\ell = |i - j|}^{\mathrm{Min}(i+j, k - i -j)} \left[2\ell+1\right] \lambda^{-2}_{\ell,+}(i,j) \nonumber\\
&=& \sum_{\ell = |i - j|}^{\mathrm{Min}(i+j, k - i -j)} \left(\frac{q^{\ell+1/2} - q^{-\ell-1/2}}{q^{1/2}- q^{-1/2}}\right) q^{-i(i+1) - j(j+1)  + \ell(\ell+1)}\nonumber\\
 &=& \left(\frac{q^{-i(i+1)-j(j+1)}}{q^{1/2}- q^{-1/2}}\right)\sum_{\ell = |i - j|}^{\mathrm{Min}(i+j, k - i -j)} \left( q^{(\ell+1)^2-1/2} - q^{\ell^2-1/2}\right) \nonumber\\
 &=&  \left(\frac{q^{-i(i+1)-j(j+1)}}{q^{1/2}- q^{-1/2}}\right) \left( q^{(\mathrm{Min}(i+j,k-i-j)+1)^2-1/2} - q^{(i-j)^2-1/2}\right) \nonumber\\
 &=& \left(\frac{q^{2ij + i + j +1/2}- q^{-2ij - i - j -1/2}}{q^{1/2}- q^{-1/2}}\right)\nonumber\\
 &=& \frac{\sin\left(\frac{\pi (2i+1)(2j+1)}{k+2}\right)}{\sin\left(\frac{\pi}{k+2}\right)}
\eeqn
which agrees with known results for the $\cS$ matrix of the $SU(2)_k$ WZW theory. (In the first line above we have used the formula
\beq
a_{0,l}\left(\begin{matrix} j_1 & j_2 \\ j_3 & j_4 \end{matrix}\right) = (-1)^{j_1+j_3-l}\sqrt{\frac{[2l+1]}{[2j_2+1][2j_3+1]}}\delta_{j_1,j_2}\delta_{j_3,j_4}.
\eeq

With these facts, we are now in a position to evaluate the Whitehead link invariant
\beqn
C_{5^2_1} (j_1, j_2) &= & [2j_1+1]^2[2j_2+1]\sum_{\bs{m, n, p}}  {\lambda}^{-1}_{ p_1,-}(j_1,j_2) \lambda_{p_2,+}(j_1,j_2) {\lambda}^{-1}_{n_1,+}(j_1,j_2){\lambda}^{-1}_{ m_1,-}(j_1,j_2) \lambda_{m_2,+}(j_1,j_2)\nonumber\\
&\times &a_{(\bs 0, \bs p)} \left(\begin{matrix} j_1 &  j_1 \\ j_2 &  j_2 \\  j_1 & j_1 \end{matrix}\right) a_{(\bs n, \bs p)} \left(\begin{matrix} j_1 &  j_2 \\ j_1 &  j_1 \\  j_2 & j_1 \end{matrix}\right)a_{(\bs n, \bs m)} \left(\begin{matrix} j_1 &  j_2 \\ j_1 &  j_1 \\  j_2 & j_1 \end{matrix}\right)a_{(\bs 0, \bs m)} \left(\begin{matrix} j_1 &  j_1 \\ j_2 &  j_2 \\  j_1 & j_1 \end{matrix}\right)
\eeqn

Similarly, we can also use the same techniques to evaluate the link invariant corresponding to the Borromean rings (figure \ref{fig:fig19.pdf}). In this case, we find

\myfig{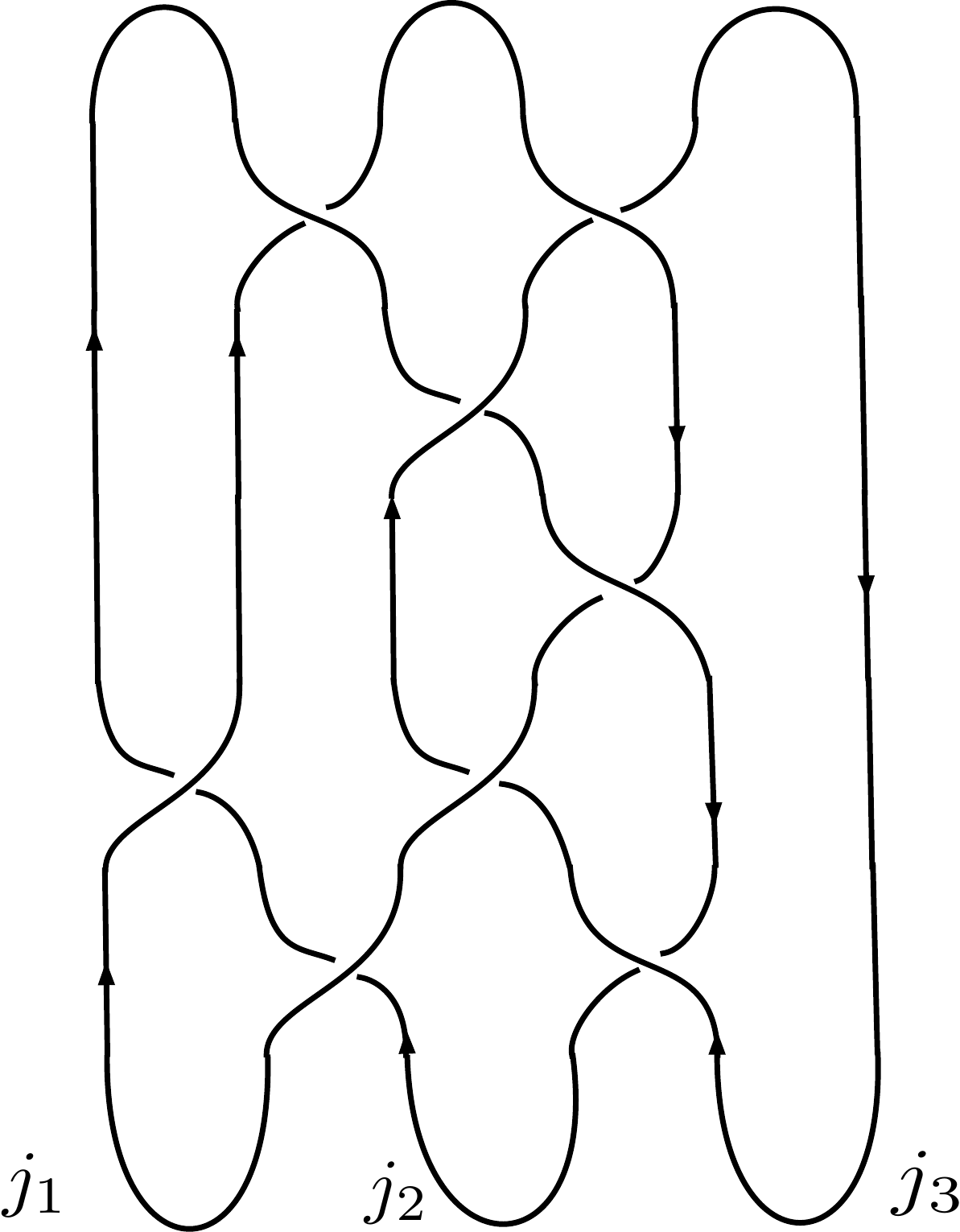}{5}{\small{\textsf{A plait representation for $6^3_2$, Borromean rings.}}}

\beqn
C_{6^3_2} (j_1, j_2, j_3) &= & \left\langle \phi_{\bs 0} (\bar{j}_1, j_1, \bar{j_2}, j_2, \bar{j_3}, j_3) \right| B_2B_4^{-1}B_1 B_3B_4^{-1}B_3B_2^{-1}B_4^{-1} \left| \phi_{\bs 0}(j_2,\bar{j_2}, j_1,\bar{j}_1, j_3,\bar{j}_3)\right\rangle\nonumber\\
&=& [2j_1+1][2j_2+1][2j_3+1]\sum_{\bs{l},\bs{m},\bs{n},\bs{p},\bs{q}} \lambda_{l_1,-}(j_1,j_2)\lambda_{l_2,-}(j_1,j_3)\lambda_{m_1,-}^{-1}(j_2,j_3)\lambda^{-1}_{n_2,+}(j_1,j_2)\nonumber\\
&\times & \lambda_{p_0,+}(j_1,j_2)\lambda_{p_1,-}^{-1}(j_1,j_3)\lambda_{q_1,-}^{-1}(j_1,j_2)\lambda_{q_2,-}(j_2,j_3) \nonumber\\
&\times & a_{(\bs 0, \bs l)} \left(\begin{matrix} j_2 &  j_2 \\ j_1 &  j_1 \\  j_3 & j_3 \end{matrix}\right)a_{(\bs m, \bs l)} \left(\begin{matrix} j_2 &  j_1 \\ j_2 &  j_3 \\  j_1 & j_3 \end{matrix}\right)a_{(\bs m, \bs n)} \left(\begin{matrix} j_2 &  j_1 \\ j_3 &  j_2 \\  j_1 & j_3 \end{matrix}\right)\nonumber\\
&\times & a_{(\bs p, \bs n)} \left(\begin{matrix} j_2 &  j_1 \\ j_3 &  j_1 \\  j_2 & j_3 \end{matrix}\right)a_{(\bs p, \bs q)} \left(\begin{matrix} j_1 &  j_2 \\ j_1 &  j_3 \\  j_2 & j_3 \end{matrix}\right) a_{(\bs 0, \bs q)} \left(\begin{matrix} j_1 & j_1 \\ j_2 & j_2 \\ j_3 & j_3 \end{matrix}\right)
\eeqn

\section{Relative entropies of links}
As mentioned in the body of the paper, the entanglement spectrum of a given link reduced on one or more of its components is a coarse measure of its topological properties.  This is well illustrated particularly by the $2^2_1+2^2_1$ link depicted in Figure \ref{fig:fig11.pdf}.  Despite $L_1$ and $L_2$ playing very different roles in the link, the reduced density matrices $\rho_{L_1}(2^2_1+2^2_1)$ and $\rho_{L_2}(2^2_1+2^2_1)$ have identical spectrum.  Additionally this spectrum is also found in a completely different link, $6^3_3$, depicted in Figure \ref{fig:fig12.pdf} reduced on one of its components.  In these cases we expect relative entropy to provide a basis independent method to distinguish reduced density matrices.  The relative entropy, $\relent{\rho}{\sigma}$ is defined as:
\begin{equation}
\relent{\rho}{\sigma}=\mTr\left(\rho\ln\rho\right)-\mTr\left(\rho\ln\sigma\right).
\end{equation}
In this appendix we outline the two calculations of the relative entropy from the main text.
\subsection{$2^2_1+2^2_1$}
Let us begin with the two different ways of reducing the $2^2_1+2^2_1$ state:  %The density matrix associated to this link is
%\begin{equation}
%\rho(2^2_1+2^2_1)=n^{-1}\sum_{ijk}\sum_{lmn}\frac{S_{ij}S_{kj}}{S_{0j}}\frac{S^*_{lm}S^*_{nm}}{S_{0m}^*}|ijk\rangle\langle lmn|
%\end{equation}
%with normalization
%\begin{equation}
%n=\sum_j\frac1{\abs{S_{0j}}^2}.
%\end{equation}
%We have two choices for reducing this state: 
we can either trace over $L_2$ and $L_3$ or we can trace over $L_1$ and $L_3$.  We are interested in calculating $\relent{\rho_{L_1}}{\rho_{L_2}}$.  Since $S_{EE}(\rho_{L_1|L_2,L_3})$ is known, what remains is the calculation of $\mTr\left(\rho_{L_1}\ln\rho_{L_2}\right)$.  Tracing over $L_2, L_3$ gives the reduced density matrix
\begin{equation}\label{red1sumHopf}
\rho_{L_1}(2^2_1+2^2_1)=n^{-1}\sum_{j}\sum_{ik}\frac{1}{\abs{\cS_{0j}}^2}\cS_{ij}\cS_{kj}|i \rangle\langle k|.
\end{equation}
with normalization $n=\displaystyle\sum_j\frac1{\abs{\cS_{0j}}^2}$. 
%We note now for future convenience that this matrix
%This matrix is diagonalized by the unitary change of basis
% $|ji\rangle=\displaystyle\sum_{\tilde i\tilde j}\delta_{j\tilde j}S^*_{i\tilde i}|\tilde j\tilde i\rangle$:
%\begin{equation}
%\rho_{23}(2^2_1+2^2_1)=n^{-1}\sum_{\tilde j}\sum_{\tilde i}\frac{1}{\abs{S_{0j}}^2}\delta_{\tilde i\tilde j}|\tilde j\tilde i\rangle\langle \tilde j\tilde i|.
%\end{equation}
%In this basis 
%under which we can easily evaluate the logarithm.  Moving back to the original basis: %by undoing the unitary transformation:
%\begin{equation}
%\ln\rho_{L_1|L_2,L_3}(2^2_1+2^2_1)=\sum_j\sum_{ik}\ln\left(p_j\right)S_{ij}S_{kj}^*|ji\rangle\langle jk|.
%\end{equation}
 Now we look at the reduced state from tracing over $L_1, L_3$:
\begin{equation}
\rho_{L_2}(2^2_1+2^2_1)=n^{-1} \sum_j\frac{1}{\abs{\cS_{0j}}^2}| j \rangle\langle j |.
\end{equation}
These expressions can more simply be written in terms of the orthonormal basis $|\hat{j}\rangle = \sum_i\cS_{ij}|i\rangle$.
%It is simple to see the unitary transformation that diagonalizes this matrix and the logarithm is easily extracted.  In the original basis we have
%\begin{equation}
%\ln\rho_{13}=\sum_j\sum_{ik}\sum_{ln}\ln\left(p_j\right)\cS_{ij}\cS_{kj}\cS_{lj}^*\cS_{nj}^*|ik\rangle\langle ln|.
%\end{equation}
From there, it is a simple matter to compute %the relative entropy between these two states:
%\begin{align}
%\mTr\left(\rho_{23}\ln\rho_{13}\right)=&n^{-1}\sum_j\frac{1}{\abs{\cS_{0j}}^2}\ln(p_j)\abs{\cS_{jj}}^2.
%\end{align}
\beqn
\mTr\left(\rho_{L_1}\ln\rho_{L_1}\right)=
\sum_i p_i\ln p_i,\qquad
\mTr\left(\rho_{L_1}\ln\rho_{L_2}\right)=
\sum_{i,j} p_i|\cS_{ij}|^2\ln p_j
\eeqn
where we recall $p_j=\frac{d_j^{-2}}{\sum_i d_i^{-2}}$. The relative entropy between these two states is thus
\begin{equation}
%\relent{\rho_{L_1|L_2,L_3}(2^2_1+2^2_1)}{\rho_{L_2|L_1,L_3}(2^2_1+2^2_1)}
\relent{\rho_{L_1}}{\rho_{L_2}}=\sum_i p_i \left(\mathrm{ln}\,p_i - \sum_j |\cS_{ij}|^2\mathrm{ln}\, p_j\right)
\end{equation}
It is straightforward to check that the relative entropy we obtained above is manifestly positive.\footnote{One could also also compute relative entropies of two component states obtained by tracing out one link. In some situations, this leads to infinite answers.}
\subsection{$6^3_3$ vs. $2^2_1+2^2_1$}
Now we comment on the spectrum of $6^3_3$ and $2^2_1+2^2_1$ reduced on to a single component.  In this case, it is useful to reduce $6^3_3$ on $L_2$ yielding a reduced density matrix
\begin{equation}
\rho_{L_2}(6^3_3)= n^{-1}\sum_{j}\frac{1}{\abs{\cS_{0j}}^2}|\tilde{j} \rangle\langle \tilde{j} |.
\end{equation}
with $n$ the same as before, and we have introduced the orthonormal basis $|\tilde j\rangle\equiv\sum_m e^{-2\pi ih_m}\cS_{mj}|m\rangle$. %, this can be simplified to
%\beq
%\rho_{L_1|L_2,L_3}(6^3_3)=\sum_j p_j |\tilde{j}\tilde{j}\rangle\langle\tilde{j}\tilde{j}|.
%\eeq
 Now let us compare this to $2^2_1+2^2_1$ reduced on $L_2$ by computing $\relent{\rho_{L_2}(6^3_3)}{\rho_{L_2}(2^2_1+2^2_1)}$. We find
\beq
\relent{\rho_{L_2}(6^3_3)}{\rho_{L_2}(2^2_1+2^2_1)}= \sum_i p_i \left(\mathrm{ln}\,p_i - \sum_j |\cS_{ij}|^2\mathrm{ln}\, p_j\right).
\eeq
% The thing to compute here is $\mTr\left(\rho_{L_1|L_2,L_3}(6^3_3)\ln\rho_{L_1|L_2,L_3}(2^2_1+2^2_1)\right)$.  From \eqref{red1sumHopf} we can see easily see the unitary basis that diagonalizes $\rho_{L_1|L_2,L_3}(2^2_1+2^2_1)$ and read off the logarithm.  In the original basis we have
%\begin{equation}
%\ln\rho_{L_1|L_2,L_3}(2^2_1+2^2_1)=\sum_j\sum_{ik}\ln\left(p_j\right)\cS_{ij}\cS_{kj}^*|ji\rangle\langle jk|.
%\end{equation}
%From here the matrix multiplication and trace is simple to evaluate and we find a relative entropy
%\begin{equation}
%\relent{\rho_{L_1|L_2,L_3}(6^3_3)}{\rho_{L_1|L_2,L_3}(2^2_1+2^2_1)}=\sum_{il}\left(\delta_{il}-\abs{\cS_{il}}^2\abs{\sum_{j}e^{2\pi ih_j}\cS_{ji}^*\cS_{jl}}^2\right)p_l\ln(p_i).
%\end{equation}

\subsection{Distinguishability of two component links}
%Although all of the two component links are easily distinguished by their entanglement entropy, let us comment relative entropy of two component links.  
For three component links the relative entropy is a useful way of comparing links with similar entanglement spectrum.  For all of the two component links we considered above, their entanglement spectrum was enough to distinguish different links.  A natural question one might want to consider in this context, however is whether the entanglement spectrum can characterize how different two links are; for simplicity let us consider how different a given link is from some fiducial simple link, for example the Hopf link, $2^2_1$. The natural tool to address this question is the relative entropy of links reduced on one of their components.  In fact this question is particularly simple to address and the answer is that the distinguishability of the link is entirely encoded in its entanglement spectrum.  To see this we note that $2^2_1$ is the maximally mixed state:
%The question we wish to address is whether the relative entropy can give us a novel gauge of ``complexity of a link" by comparing it to the Hopf link.  That is to say, if we reduce the Hopf link by tracing over a subspace and compute the relative entropy with respect to another reduced link does the answer give us anything novel?  The short answer is no: in this case the relative entropy is essentially characterized by the complex link's entanglement entropy, or more precisely, its entanglement spectrum.  To see this we note that the reduced Hopf link is the maximally mixed state:
\begin{equation}
\rho_{L_2|L_1}(2^2_1)=\frac{1}{\text{dim}\mathcal H_{T_2}}\sum_i |i\rangle\langle i|.
\end{equation}
Because of this, for any diagonalizable density matrix, $\tilde\rho_{L_2|L_1}$, on $\mathcal H_{T_2}$ obtained by reducing a two component link on its second component,\footnote{In fact this argument works for any $n$ component link reduced on $n-1$ of its components $\tilde\rho_{L_1\ldots L_{n-1}|L_n}$.} we can simultaneously diagonalize $\tilde\rho_{L_2|L_1}$ and $\rho_{L_2|L_1}(2^2_1)$.  Let the spectrum of $\tilde\rho_{L_2|L_1}$ be $\{\tilde p_i\}_{i\in\text{span}(\mathcal H_{T_2})}$.  Then it is a simple exercise to show that
\begin{equation}
\relent{\tilde\rho_{L_2|L_1}}{\rho_{L_2|L_1}(2^2_1)}=-S(\tilde\rho)-\sum_i\tilde p_i\ln\left(\frac{1}{\text{dim}\mathcal H_{T_2}}\right)=\ln\left(\text{dim}\mathcal H_{T_2}\right)-S(\tilde\rho)
\end{equation}
where we used $\sum_i\tilde p_i=1$.  %Similarly
%\begin{equation}
%S(\rho_1(2^2_1)||\tilde\rho)=-\ln\left(\text{dim}\mathcal H_{T_2}\right)-\frac{1}{\text{dim}\mathcal H_{T_2}}\sum_i\ln\tilde p_i.
%\end{equation}
Therefore the distinguishability of a two component link from the Hopf link amounts to only knowing that link's entanglement spectrum.
% which is typically extracted as a bonus during the entanglement entropy calculation.  Note that this is a special property of the Hopf link; comparing generic two component links is more non-trivial.

%\bibliographystyle{uiuchept}
%\bibliography{Links}
\providecommand{\href}[2]{#2}\begingroup\raggedright\endgroup

\end{document}